\documentclass[journal=ancac3,manuscript=article]{achemso}

\usepackage[version=4]{mhchem} 
\usepackage{xcolor}

\author{Wangwei Dong}
\altaffiliation{These authors contributed equally.}
\author{Zezhou Liu}
\altaffiliation{These authors contributed equally.}
\affiliation[McGill University]
{Department of Physics, McGill University, Montr\'eal, Qu\'ebec, Canada, H3A 2T8}
\author{Ruiyao Liu}
\affiliation[UCSB University]
{Department of Physics, University of California, Santa Barbara, CA 93106}
\author{Deborah Kuchnir Fygenson}
\email{fygenson@ucsb.edu}
\affiliation[UCSB University]
{Department of Physics, University of California, Santa Barbara, CA 93106}
\alsoaffiliation[UCSB]
{interdisciplinary program in Quantitative Biology, University of California, Santa Barbara, CA 93106}
\author{Walter Reisner}
\email{walter.reisner@mcgill.ca}
\affiliation[McGill University]
{Department of Physics, McGill University, Montr\'eal, Qu\'ebec, Canada, H3A 2T8}

\title[]
  {DNA Nanostructures Characterized via Dual Nanopore Resensing}


\begin{document}

\begin{tocentry}
\centering
\includegraphics[height=4.45cm]{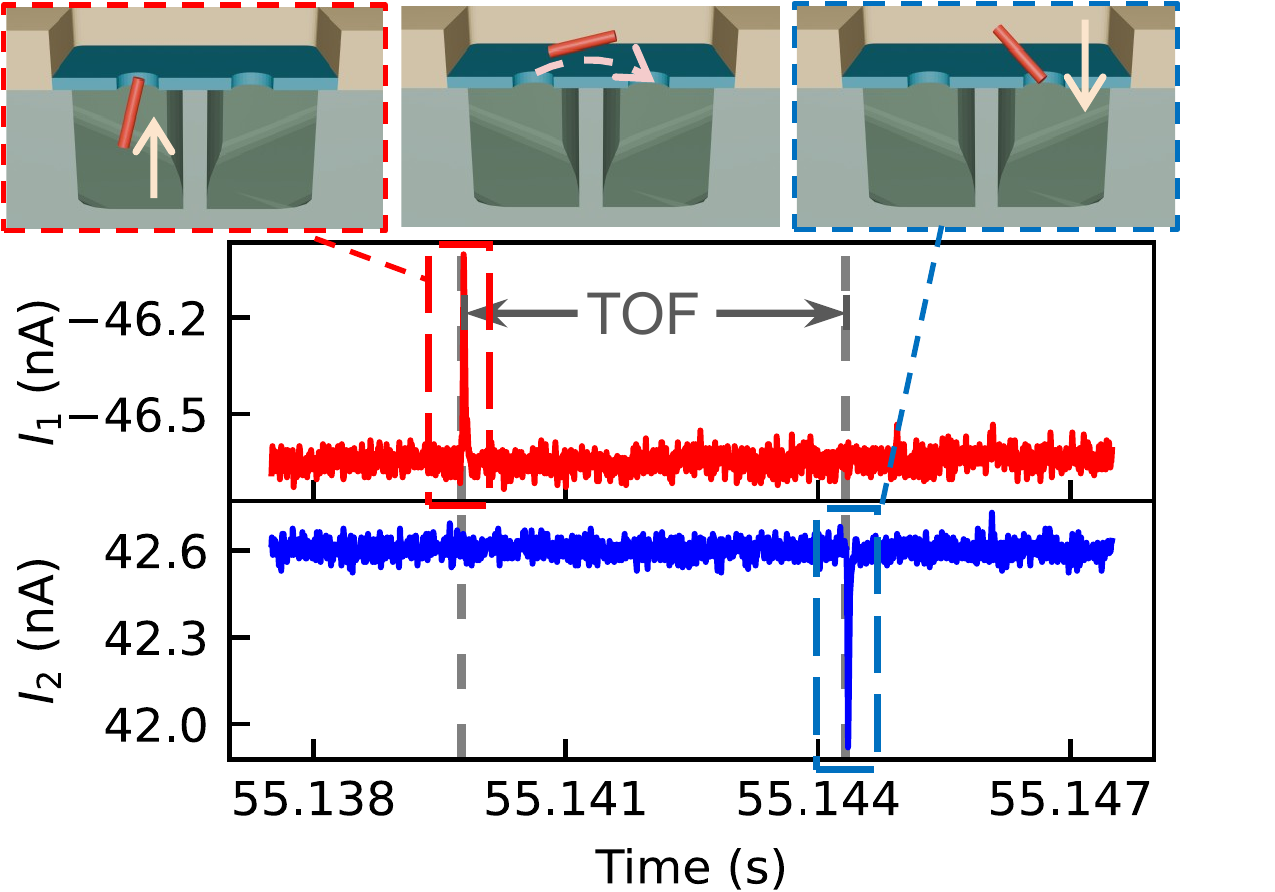}

\end{tocentry}

\begin{abstract}
  DNA nanotechnology uses predictable interactions of nucleic acids to precisely engineer complex nanostructures. Characterizing these self-assembled structures at the single-structure level is crucial for validating their design and functionality. Nanopore sensing is a promising technique for this purpose as it is label-free, solution-based and high-throughput.  Here, we present a device that incorporates dynamic feedback to control the translocation of DNA origami structures through and between two nanopores. We observe multiple translocations of the same molecule through the two distinct nanopores as well as measure its time-of-flight between the pores.  We use machine learning classification methods in tandem with classical analysis of dwell-time/blockade distributions to analyze the complex multi-translocation events generated by different nanostructures.  With this approach, we demonstrate the ability to distinguish DNA nanostructures of different lengths and/or small structural differences, all of which are difficult to detect using conventional, single-nanopore sensing. In addition, we develop a finite element diffusion model of the time-of-flight process and estimate nanostructure size. This work establishes the dual nanopore device as a powerful tool for DNA nanostructure characterization.
\end{abstract}

\section{Introduction}
The specific base-pairing and predictable interactions of DNA make DNA nanotechnology a powerful approach for creating custom-designed nanostructures. 
This technology uses DNA strands as building blocks that self-assemble into 2-dimensional (2D) and 3-dimensional (3D) nanostructures with precision and programmability \cite{Seeman2017}. 
Taking advantage of the outstanding stability and versatility of DNA, DNA nanotechnology is proving useful in a wide variety of applications in biophysics, drug delivery, biosensing, tissue regeneration, therapeutics, material engineering, \textit{etc.} \cite{Seeman2017,Ma2021,Hu2019}.  When designing and optimizing DNA nanostructures for such applications, it is essential to determine whether, and to what extent, the nanostructures form as designed \cite{Mathur2017,Techniques,Knappe2023}.  

Three approaches are commonly used to characterize DNA nanostructures: gel electrophoresis, atomic force microscopy (AFM) and cryo-electron microscopy (CEM).
Gel electrophoresis is a simple, inexpensive and non-destructive characterization method, but is limited by low sensitivity and low resolution.\cite{Zhu2018}
AFM enables direct visualization of individual DNA nanostructures with high resolution. 
However, AFM requires surface immobilization that tends to distort or wholly disrupt three-dimensional structures \cite{Platnich2020}.
CEM provides composite visualization with minimal perturbation, but is expensive, technically challenging, and low-throughput.\cite{mi14010118}
Therefore, new approaches are needed for high-resolution, solution-based characterization of DNA nanostructures.

In the nanopore approach, an applied voltage drives ions through a nanoscale hole in an electrically insulating membrane between two reservoirs filled with electrolyte solution.  
When an analyte moves through the pore, it changes the electrical resistance of the circuit, resulting in a detectable change in current, commonly called a `conductance blockade'. 
Information about the analyte, such as its volume, charge and shape, can be inferred from features like amplitude and duration of the conductance blockade \cite{Miles2013,Xue2020}. 
This signal has been used to identify nanostructure tags along DNA strands \cite{Plesa2015,Beamish2017,Chen2023}, probe the local rigidity of nanostructures \cite{He2023}, and distinguish different DNA nanostructures \cite{DNAstar,cube-ring,four-brick}. 

To date, nanopore-based differentiation between nanostructures has largely relied on their distinct conductance blockade distributions. 
For example, DNA stars with different numbers of arms were classified based on significant differences in their current blockade profiles.\cite{DNAstar} 
DNA cubes and RNA rings also yielded clearly distinguishable current blockades.\cite{cube-ring}  
Similarly, nanostructures composed of different numbers of pre-assembled DNA origami building blocks showed little to no overlap in their blockade distributions.\cite{four-brick}
While these works successfully demonstrate the utility of nanopore sensing for structural classification, the studies rely on gross structural differences to produce distinguishable current blockade levels. 
Accurate classification of nanostructures whose structural differences are so subtle that their current blockade distributions overlap has not been demonstrated.  
In addition, quantitative estimation of DNA nanostructure dimensions through nanopore sensing remains unexplored.

Here we show that a dual-nanopore platform operated with active logic enabling multiple translocations of the same analyte can classify and characterize closely related DNA nanostructures. We benchmark our technology with  12-helix hollow cylinders synthesized \textit{via} DNA origami.\cite{seed}   These nanostructures are used in the field of DNA nanotechnology as seeds for nucleating DNA nanotubes, a major sub-class of DNA nanostructure.\cite{DNANanotubeReview}  Nanotubes possess a hollow geometry with controllable stiffness and surface interactions that makes them ideal for a range of applications \cite{DNANanotubeReview}, including as linear templates guiding the assembly of other nanoscale components (\textit{e.g.} proteins \cite{ProteinOrgNT}, and Au nanoparticles \cite{TubeAuNP}); as tracks for natural \cite{NTTrack} and synthetic \cite{SynMotorNT} molecular motors; as vehicles for drug delivery \cite{drugdelivery}; and, as building blocks to produce a wide range of novel functional materials (\textit{e.g.} liquid crystals that align membrane proteins for NMR \cite{NTLCNMR} and substrates for tissue assembly \cite{NTTissue}). For this study, the seeds were intentionally chosen to be structurally similar in order to challenge the sensing methodology.  

Our dual pore platform permits independent voltage biasing and acquisition of the trans-pore current at each pore.  
This allows pore biasing to be adjusted in response to current feedback at each pore, enabling complex control modes such as the ability to drive a single nanostructure back and forth through each pore and between the pores.\cite{small2018,small2019,small2020}   
We show that, despite major overlaps of conductance blockade and dwell-time distributions for single-pore translocation events, comprehensive results based on multiple translocations of the same structure through two pores can differentiate between closely related nanostructures with an accuracy of 83\% using machine learning classification methods. 
Moreover, the ability to obtain multiple translocations of a single nanostructure enables determination of the event-to-event heterogeneity arising purely from intrinsic variability in translocation physics.  
This helps deconvolve variability arising from translocation physics from variability arising from chemical and structural differences between the analytes.  
Finally, the ability to translocate a single structure between two pores enables acquisition of the structure's time-of-flight (TOF), allowing quantification of its diffusional dynamics.  
Using a finite element diffusion model of inter-pore dynamics, we use the TOF measurements to estimate the diffusion constant, from which we extract quantitative estimates of a cylindrical nanostructure's length.

\section{Results and Discussion}

\subsection{Dual nanopore resensing platform}

\begin{figure}
    \centering
    \includegraphics[width=0.9\linewidth]{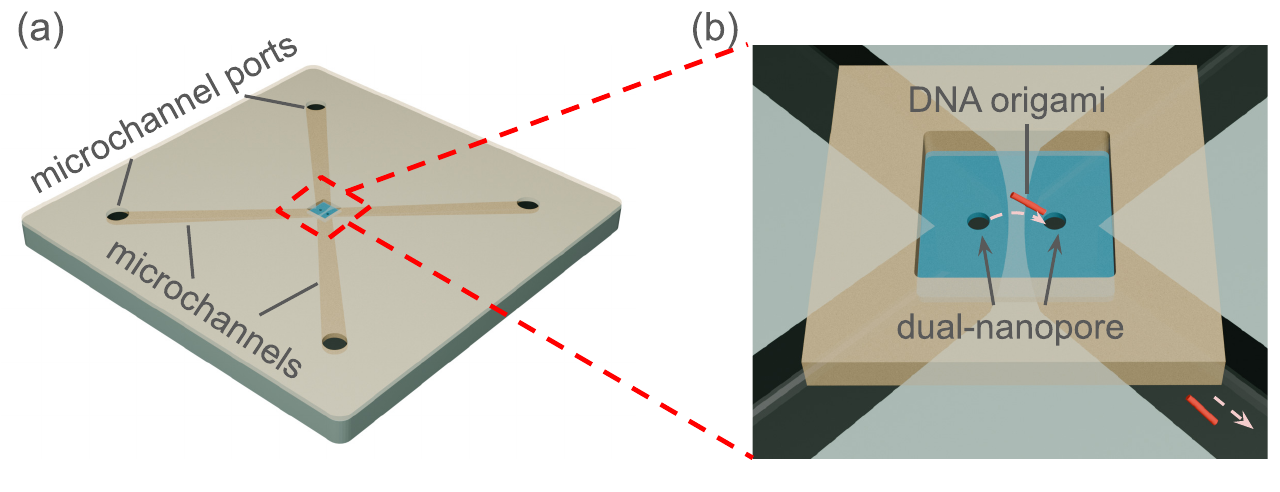} 
  \caption{Dual-nanopore chip schematic. 
  (a) A 3D schematic of the chip. 
  (b) Zoomed-in view of the dual-nanopore structure. 
  Two nanopores are positioned at the tips of the two ``V''-shaped microchannels. 
  Each nanopore thereby interfaces the common chamber above the membrane to a separate fluidic channel that can be independently addressed.}
  \label{chip}
\end{figure}

Our dual-nanopore chip features two microchannels interfaced to a solid-state SiN membrane (Fig.~\ref{chip}). 
Each microchannel is coupled to an individual membrane-spanning pore of $\sim 30$~nm diameter, drilled at the point of closest approach of the channels \textit{via} focused ion beam (FIB).  
The two pores feed into a common chamber, enabling the translocation of analytes from one pore to the other.  
Loading ports at the microchannel termini are used to introduce high-ionic strength buffer.  
Ag/AgCl electrodes, inserted into the ports and common chamber, interface the device to a two-channel voltage-clamp amplifier coupled to a field programmable gate array (FPGA).  
The common chamber is always grounded during device operation.

\begin{figure}
    \centering
    \includegraphics[width=1\linewidth]{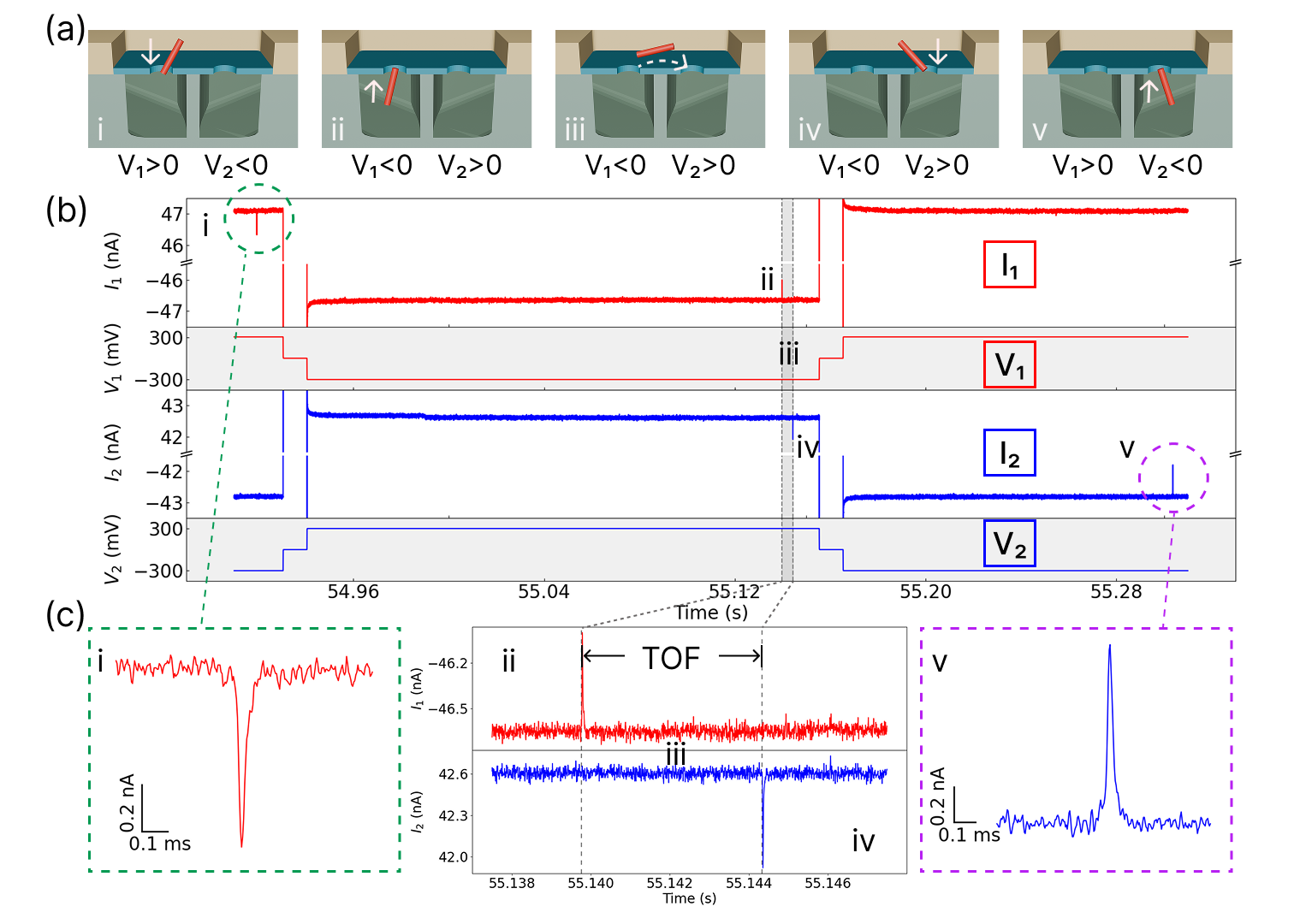} 
  \caption{resensing DNA nanostructures with active feedback control in a dual-pore device and representative signals. 
  (a) Graphical depiction of a DNA nanostructure resensing cycle. 
  A nanostructure is driven through pore 1 (left) under a positive voltage $V_1$ while the pore 2 voltage $V_2$ is held negative (i). 
  Both voltages are then reversed so the same nanostructure translocates back through pore 1 (ii) and moves toward pore 2 (iii). 
  Upon detection of nanostructure translocation through pore 2 (iv), the voltages are reversed again so the nanostructure translocates back through pore 2 (v). 
  (b) The current traces ($I_1$, $I_2$) and voltages ($V_1$, $V_2$) during a resensing cycle. (c) Close-ups of the current signals of pore 1 sensing (i) and resensing (ii), definition of time-of-flight (TOF, iii), and pore 2 sensing (iv) and resensing (v). }
  \label{algorithm}
\end{figure}

The FPGA uses current feedback from translocation events to dynamically adjust nanopore biasing. 
In particular, a decrease in pore current below a threshold value, indicating a successful translocation, triggers a change in pore biasing. 
This active feedback control enables our dual pore resensing algorithm.  
Figure \ref{algorithm}a illustrates the steps in a resensing cycle. 
By convention, we refer to the left pore/channel as pore/channel 1 and the right pore/channel as pore/channel 2. 
After DNA origami nanostructures are loaded into the common chamber, the voltage of channel 1 ($V_1$) is set positive and the voltage of channel 2 ($V_2$) is set negative, so as to direct the nanostructures toward pore 1. 
When the translocation of a nanostructure through pore 1 is detected from a drop in current $I_1$ (pore 1 sensing, step (i)), both voltages are reversed. 
The nanostructure that just entered channel 1 is then driven back through pore 1 (pore 1 resensing, step (ii)). 
The same nanostructure is simultaneously directed toward pore 2 (step (iii)) until its translocation through pore 2 is detected (pore 2 sensing, step (iv)). 
The time between pore 1 resensing and pore 2 sensing ({\em i.e.}, the duration of step (iii)) is defined as the time-of-flight (TOF). 
Upon detecting translocation through pore 2, the voltages are again reversed to drive the nanostructure back into the common chamber and towards pore 1 (pore 2 resensing, step (v)) where it has the possibility of translocating through pore 1 again, starting another resensing cycle. 
Figure \ref{algorithm}b shows an example of the current and voltage traces during a resensing cycle. 
Upon detection of translocations from the common chamber into the channels, the voltages are held unchanged for 11\,ms (steps (i), (iv)) to drive the nanostructure deep enough into the channel to prevent its immediate retranslocation upon field reversal.  This prevents translocations from occurring during the capacitive transient when the current is unstable.
This is followed by turning off both voltages ($V_1$=$V_2$=0) for a 10\,ms waiting time, then turning the voltages on again with the opposite polarity.  
Figure \ref{algorithm}c shows labeled and zoomed-in translocation signals corresponding to each step.

\subsection{DNA nanostructures}
\begin{figure}
    \centering
    \includegraphics[width=0.9\linewidth]{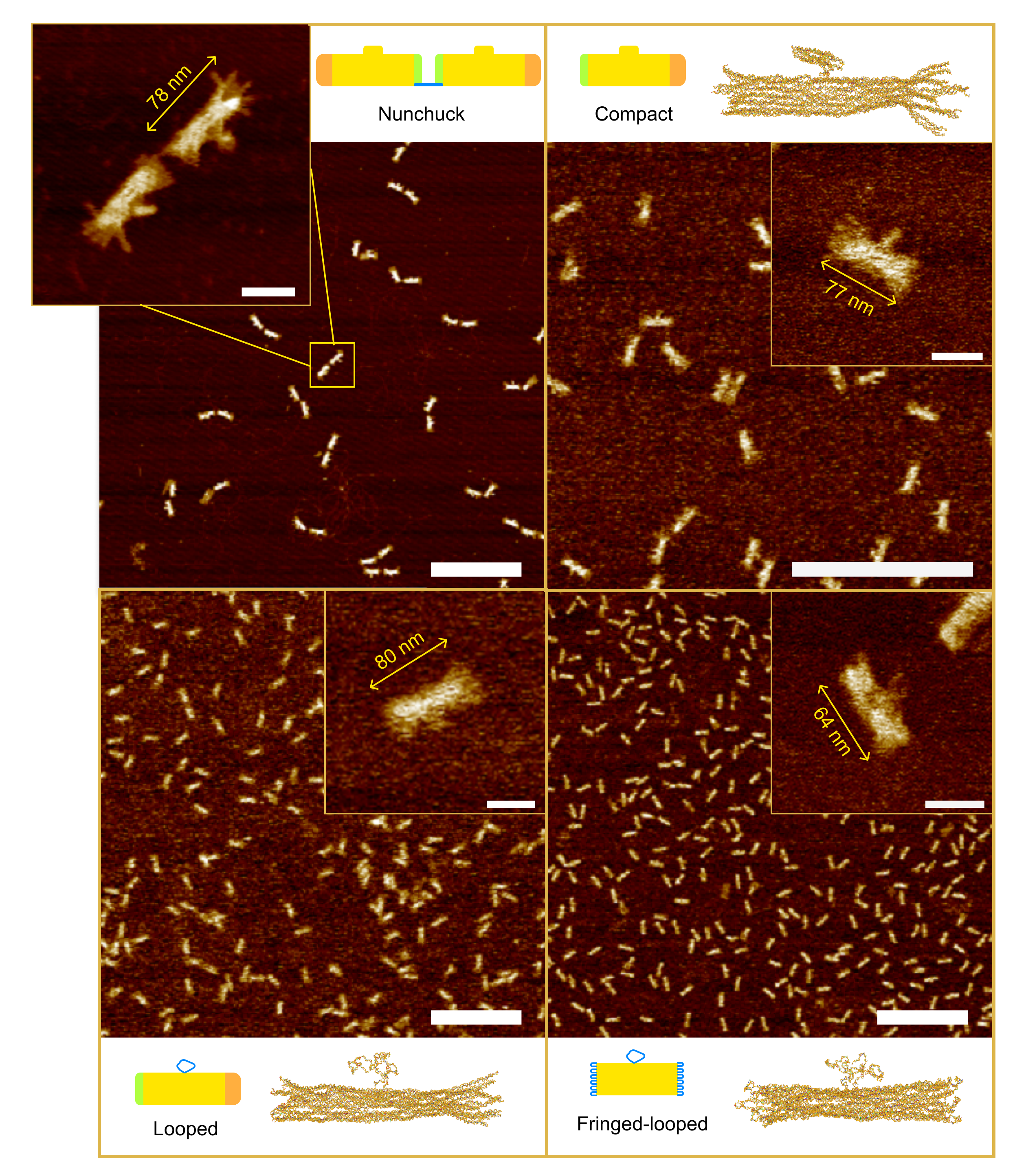} 
  \caption{AFM images of DNA nanostructures used in the study with corresponding simplified and detailed schematics. 
  Clockwise from the upper left are DNA nunchuck seeds, compact seed monomers, fringed-looped seeds, and looped seeds. 
  The scale bars represent 400~nm in the main images and 40\,nm in the insets. 
  Nunchuck monomers, compact seeds and looped seeds are all comparable in length (77-80\,nm), while fringed-looped seeds are about 12\,nm shorter. 
  In the simplified schematics, regions of double-stranded scaffold are yellow while regions of single-stranded scaffold are blue. 
  Orange and green indicate the presence of adapter tiles and blocker strands, respectively. 
  The dsDNA that links two compact seed monomers to form a nunchuck seed is dark blue. 
  The detailed schematics positioned next to their respective simplified versions were generated using oxView\cite{oxView}. 
 }
  \label{AFM}
\end{figure}

We explore the translocation behavior of four DNA nanostructures.  
Figure~\ref{AFM} shows AFM characterization of each structure, along with a schematic depiction and oxView visualizations from oxDNA simulations~\cite{oxView,Poppleton2023}. 
Their common subunit is a DNA origami, assembled using 72 short ``staple'' strands to fold a p3024 scaffold strand~\cite{Douglas} into a 12-helix, cylindrical tube. 
Short (32 base) single-stranded loops of scaffold at both ends of every helix create a ``fringe'' of binding sites for ``adapter tiles'' that allow the structure to seed the self-assembly of tiled DNA nanotubes\cite{seed}, or for ``blocker'' and ``linker'' strands that join two seeds end-wise through one, and only one, helix\cite{cai2020}.
Another 336 bases of unused scaffold emerges from one of the helices midway along the length of cylinder.
With the addition of strands that bind to the fringes, the basic ``fringed-looped'' cylinder becomes a ``looped'' seed.
Adding nine more staples folds the 336-base loop into a short, 6-helix bundle, creating a “compact” seed monomer. 
Linking two compact seeds \textit{via} a 32\,bp double-stranded DNA (dsDNA) produces nunchuck seeds\cite{nunchuck}. 
AFM measurements show that fringed-looped structures are about 12\,nm shorter than looped seeds and compact seed monomers. 
The central 6-helix bundle can be clearly seen in the images of compact seed monomers and nunchuck seeds. 

\subsection{Single pore translocation analysis}
\begin{figure}
    \centering
    \includegraphics[width=1\linewidth]{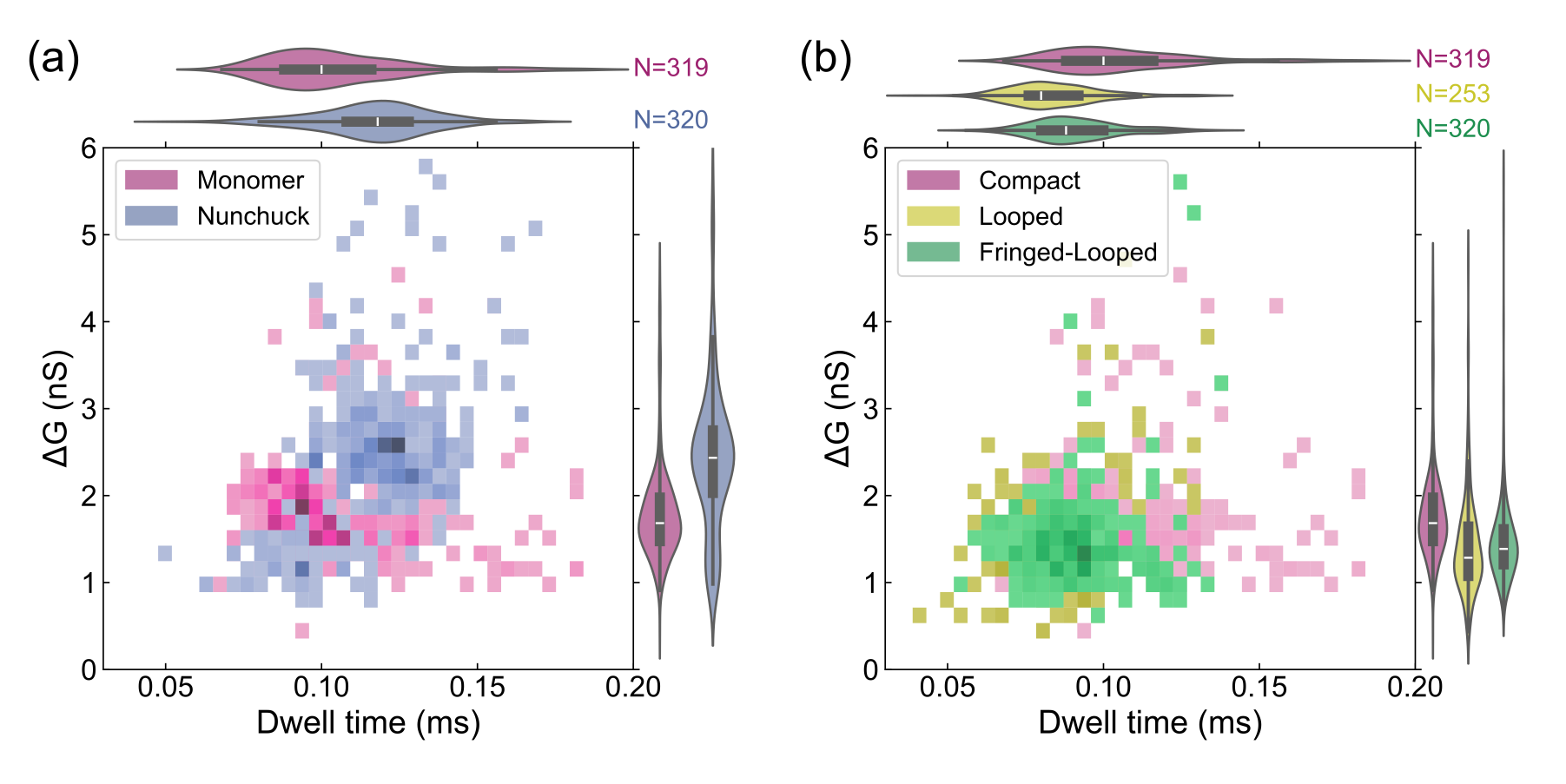} 
  \caption{Density heatmap of conductance blockade versus dwell time for single-pore translocations. 
  (a) Comparison between nunchuck and compact monomer seeds. 
  (b) Comparison of compact seed monomers, looped seeds and fringed-looped cylinders. 
  Darker shades correspond to a greater number of observations.
  On top and to the right are the corresponding violin plots (depicting area-normalized probability density functions derived from kernel density estimation, reflected about the binning axis) for each nanostructure.
The box plots inside violin plots depict first quartile - 1.5\,IQR (interquartile range), first quartile, median, third quartile and third quartile + 1.5\,IQR. The distributions for the different seeds are stacked layer-by-layer.  In (a) the nunchuck distribution is stacked on top and monomer below; in (b) the stacking order is compact, looped then fringed-looped seeds from bottom to top. Experiments were performed in 1\,M LiCl buffer at 300\,mV with a bandwidth of 30\,kHz and a pore size of 37.5\,nm.}
  \label{violin}
\end{figure}

We first present results for translocation of structures through a single-pore, specifically corresponding to the initial sensing of structure by pore 1.  These results benchmark the performance of a single-pore in our device for discriminating between different structures. Note that all events included here are followed by a pore 1 resensing event after voltage reversal.  
This ensures that analyzed events correspond to seeds that passed through, rather than just collided with, the nanopore. 
Events are characterized by conductance difference $\Delta G$ (current blockade amplitude normalized by voltage) and dwell time (event duration). 
The translocation profiles of nunchuck seeds and compact seed monomers are compared \textit{via} a density heatmap of their observed $\Delta G$ values and dwell times with corresponding marginal violin plots (Fig.~\ref{violin}a).
Compact seed monomer events produce a single cluster with a median $\Delta G$ of $1.68 \pm 0.25$\,nS and a median dwell time of $0.10\pm 0.01$\,ms. 
Nunchuck seed events produce two clusters, with a majority of events in the cluster having higher $\Delta G$ and longer dwell time; this is consistent with our expectation for the translocation of nunchuck seeds.
The lower $\Delta G$ and shorter dwell times of the minority cluster closely resemble those of compact seed monomers.
The presence of monomers in the nunchuck sample is also observed by AFM (Fig.~S3) and is likely the result of either imperfect purification or subsequent dissociation of nunchuck seeds. 
Although nunchuck translocations exhibit higher $\Delta G$ and dwell time levels than compact seeds, the heatmap distributions of the two structures partially overlap, diminishing the accuracy with which the two nanostructures can be distinguished.

The translocation profiles of compact seed monomers, looped seeds, and fringed-looped cylinders are compared in Fig.~\ref{violin}b. 
Considerable overlap is observed among the three distributions.  
We quantify this overlap \textit{via} the overlap coefficient (OVL), defined as 
\[
\mathrm{OVL}(P, Q) = \int_{-\infty}^{\infty} \min(P(x), Q(x))\, dx
\]
where $P(x)$ and $Q(x)$ are the probability density functions for the distributions being compared. 
The overlap coefficients between the distributions of the looped seeds and fringed-looped cylinders are 0.82 for $\Delta G$ and 0.78 for dwell time. 
Thus, although the fringed-looped cylinders are $\sim$12\,nm shorter, the two distributions overlap almost completely and cannot be easily distinguished, consistent with visual observation of the heatmap and violin plots.

Compact seed monomers have comparatively higher $\Delta G$ and longer dwell times than the other two seeds.  
This indicates that the excess scaffold that emerges midway along the length of the cylinder blocks more ions and slows translocation more when stapled into a 6-helix bundle than when left as flexible single-stranded loop. 
Notably, compact seeds and looped seeds are not readily differentiated by gel electrophoresis (Fig.~S1-2), highlighting the advantage of nanopore sensing in detecting subtle structural features not reflected in electrophoretic mobility.  
However, substantial overlap still exists: compact seeds have an overlap coefficient of 0.66 for $\Delta G$ and 0.72 for dwell time with fringed-looped cylinders. 
This large degree of overlap indicates that the structural differences between the nanostructures are not sufficient to differentiate them accurately using only results from single-pore translocation.

\subsection{Machine learning classification with dual pore translocations}
\begin{figure}
    \centering
    \includegraphics[width=1\linewidth]{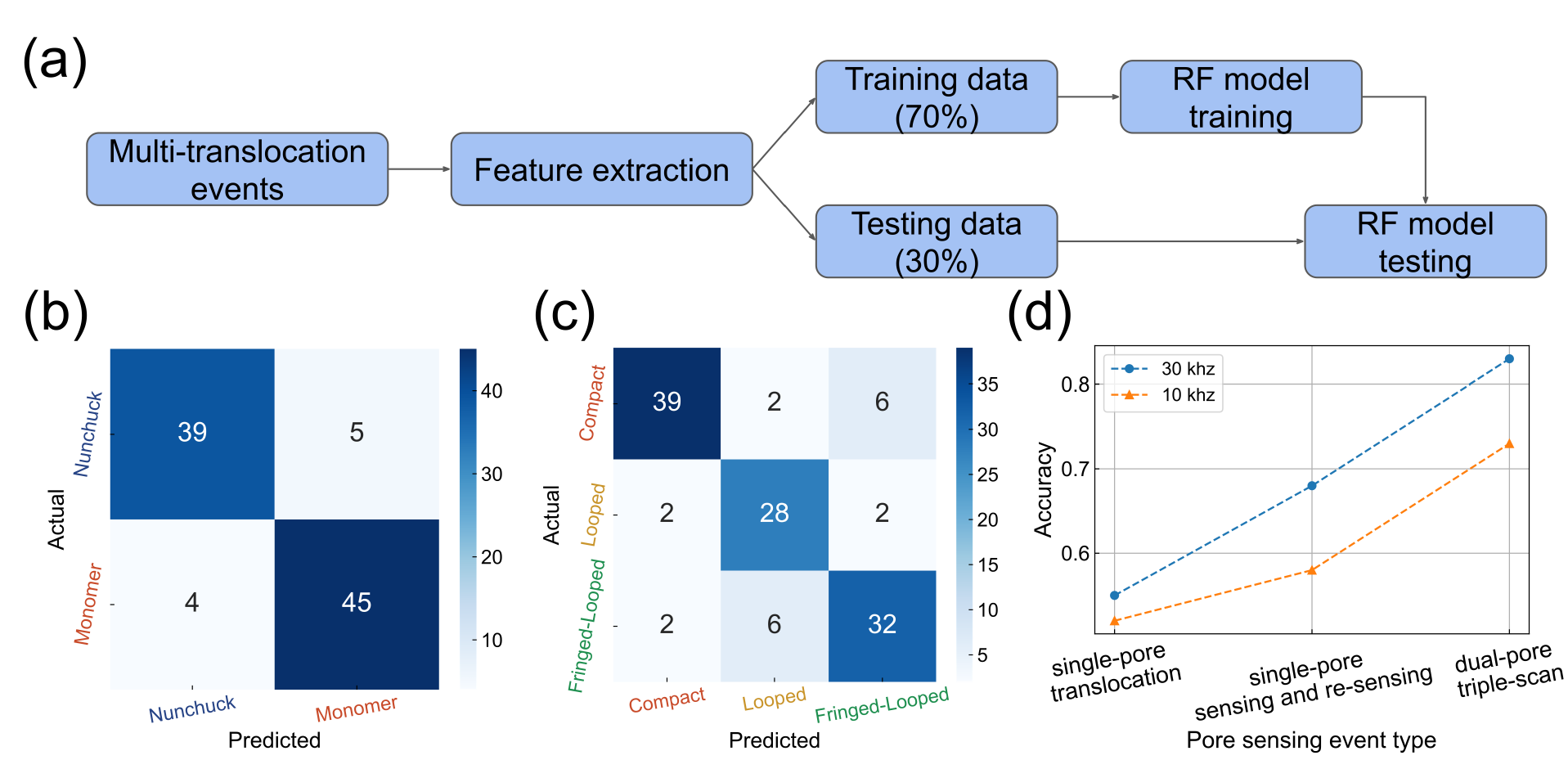} 
  \caption{Machine learning classification. 
  (a) Workflow for the classification of DNA origami seeds using multi-translocation events and Random Forest (RF) classifier.  
  Confusion matrix for classification (b) of nunchuck seeds and their compact seed monomers and (c) of compact seed monomers, looped seeds, and fringed-looped cylinders. 
  Values in the matrix indicate the number of testing multi-scan events.  
  Experiments were performed using a dual-pore chip with a pore 1 diameter of 37.5\,nm, a pore 2 diameter of 35\,nm and a bandwidth of 30\,kHz. 
  (d) Comparison of three-seed classification accuracy using a single-scan (pore 1 sensing events), double-scan (pore 1 sensing-resensing events), and triple-scan (pore 1 sensing-resensing \& pore 2 sensing events), at 10\,kHz and 30\,kHz bandwidth, respectively.
 }
  \label{ML}
\end{figure}

The integration of machine learning into nanopore sensing has helped classify nucleic acids\cite{Pandit2024,Hassan2024}, proteins\cite{Dutt2023}, and viruses\cite{Taniguchi2021}.
Here, we use machine learning to classify DNA nanostructures.
Our workflow is shown in Fig.~\ref{ML}a. 
Multi-translocation events were obtained using the dual pore resensing algorithm introduced above. 
Between each translocation event, it is possible for nanostructures to escape the dual-pore arrangement by diffusion.
In our system, ``triple-scan'' events yielded an optimal combination of number of observations ($>100$) and number of parameters for the machine learning algorithm. 
In these events, the nanostructure first exits the common chamber through pore 1 (Fig.~\ref{algorithm}(i)), returns to the common chamber through pore 1 (Fig.~\ref{algorithm}(ii)), and then leaves the common chamber through pore 2 (Fig.~\ref{algorithm}(iv)).
Using custom software adapted from the OpenNanopore \cite{opennanopore} program, we extract the conductance blockade and dwell time from each of a single nanostructure's three translocation events. 
Combined with the resensing time between the two pore 1 translocations and the time of flight (TOF) between pore 1 and pore 2, we obtain eight features for every triple-scan signal. 
We perform random forest classification to identify the nanostructures based on these features, using 70\% of the data as a training set and 30\% as a testing set.

Machine learning identification was first performed to classify nunchuck and compact seed monomers. 
The resulting confusion matrix is shown in Fig.~\ref{ML}b. 
Accuracy in machine learning refers to the proportion of correctly classified instances among all instances in the dataset. 
Here, our accuracy is as high as 0.90. 
This is not surprising because the length of a nunchuck seed is double that of a compact seed monomer and the two have well-separated clusters on the $\Delta G$-dwell time heatmap. 
For these two nanostructures, an accuracy of 0.90 can be reached even with single translocations (Fig.~S4), corroborating the effectiveness of single pores for identifying distinct structures with minor overlap in their translocation profiles.

The three types of single seeds were then classified on the basis of triple-scan events using the same workflow, leading to the confusion matrix in Fig.~\ref{ML}c. 
This identification achieved an accuracy of 0.83. 
Another performance metric, F1-score, is calculated for each individual class (defined in Section S3.2).
The compact seed had the highest F1-score (0.87, compared to 0.82 for looped seeds and 0.80 for fringed-looped cylinders), consistent with its being easily distinguished from looped seeds and fringed-looped cylinders in the density heatmap.
Since our dataset is well-balanced, the average F1-scores were equal in value to the corresponding accuracies.
The average F1-score of 0.83 from the three-seed classification was far greater than 0.33, the threshold for three-way classification \cite{Dutt2023, Taniguchi2021}. 

This high classification accuracy arises from the use of repeated translocations through the two pores, which increases the dimensionality of the data set obtained for each event, expanding the number of parameters for each measured nanostructure from two to eight.
When information from only the first translocation through pore 1 was used to classify the three seeds(\textit{i.e.}, single-pore classification), the classification accuracy was only 0.55  (Fig.~\ref{ML}(d)).
Adding conductance blockade and dwell time parameters for pore 1 resensing and the time delay between pore 1 sensing and resensing raised the accuracy to 0.68.  
Adding TOF and pore 2 conductance blockades and dwell times raised the accuracy to 0.83. 

The dual-pore approach enables high-accuracy nanostructure classification even when using a low bandwidth that leads to distortion of translocation events.  At the lower bandwidth of 10\,kHz, events shorter than the distortion point $\tau_d \approx 0.07$\,ms have suppressed amplitude and an over-estimated dwell-time \cite{distortion, distpointref} (for our four-pole Bessel filter $\tau_d\approx 0.66/\text{bandwidth}$).  This distortion effect reduced the single-pore classification accuracy to as low as 0.52.
When results from pore 1 resensing were included, the accuracy rose to 0.58.
When results from pore 2 were included, it rose to 0.73.  
Because resensing time and TOF are not affected by event distortion at low bandwidth, a dual-pore setup is  well-suited for high-accuracy nanostructure identification under conditions involving fast translocation that may challenge single pore sensing at available bandwidths.

\subsection{Molecular length estimation from diffusion dynamics modeling}
To investigate nanostructure transport through the dual pore system and compute the distributions of TOF and resensing time, we developed a three-dimensional finite element model.   
The pore-to-pore distance used in the model (674\,nm) was extracted directly from scanning electron microscopy (SEM) measurement.
The electrostatic potential distribution was computed by solving the Poisson equation,
\begin{equation}
  \nabla^2 u = 0 \label{poisson}
\end{equation}
where $u$ is the electric potential. 
The top surface of the common chamber was held at $u=0$\,mV, the channel 1 terminus was set at -300\,mV and the channel 2 terminus was set at +300\,mV.
At time $t=0$, the probability density function for the nanostructure’s center position (PDF, $p(x, y, z, t)$) was a three-dimensional, isotropic Gaussian centered at the upper edge of pore 1. 
The PDF was evolved according to the convection–diffusion equation:
\begin{equation}
  \frac{\partial p}{\partial t} = \nabla \cdot (D \nabla p - \mathbf{v} p) \label{Smoluchowski}
\end{equation}
where $D$ is the diffusion coefficient of the nanostructure and $\mathbf{v}$ is its drift velocity, which we obtain from the electric potential via:
\begin{equation}
  \mathbf{v} = -\mu \nabla u \label{velocity}
\end{equation}
where $\mu$ is the electrophoretic mobility.  
After discretization \textit{via} finite differencing, Eq.~\ref{Smoluchowski} was evolved numerically using FreeFem++ (v.4.15, freefem.org).

We model the diffusion of our nanostructures as isotropic, not explicitly taking into account the rotational dynamics of the rod-like seeds and the coupling of seed rotational to translational dynamics induced by rod anisotropy.  This assumption is justified because the rotational decorrelation time of the seeds is much shorter than their typical time-of-flight between nanopores. For a detailed justification of the isotropic model, see Section S4.3.

\begin{figure}
    \centering
    \includegraphics[width=1\linewidth]{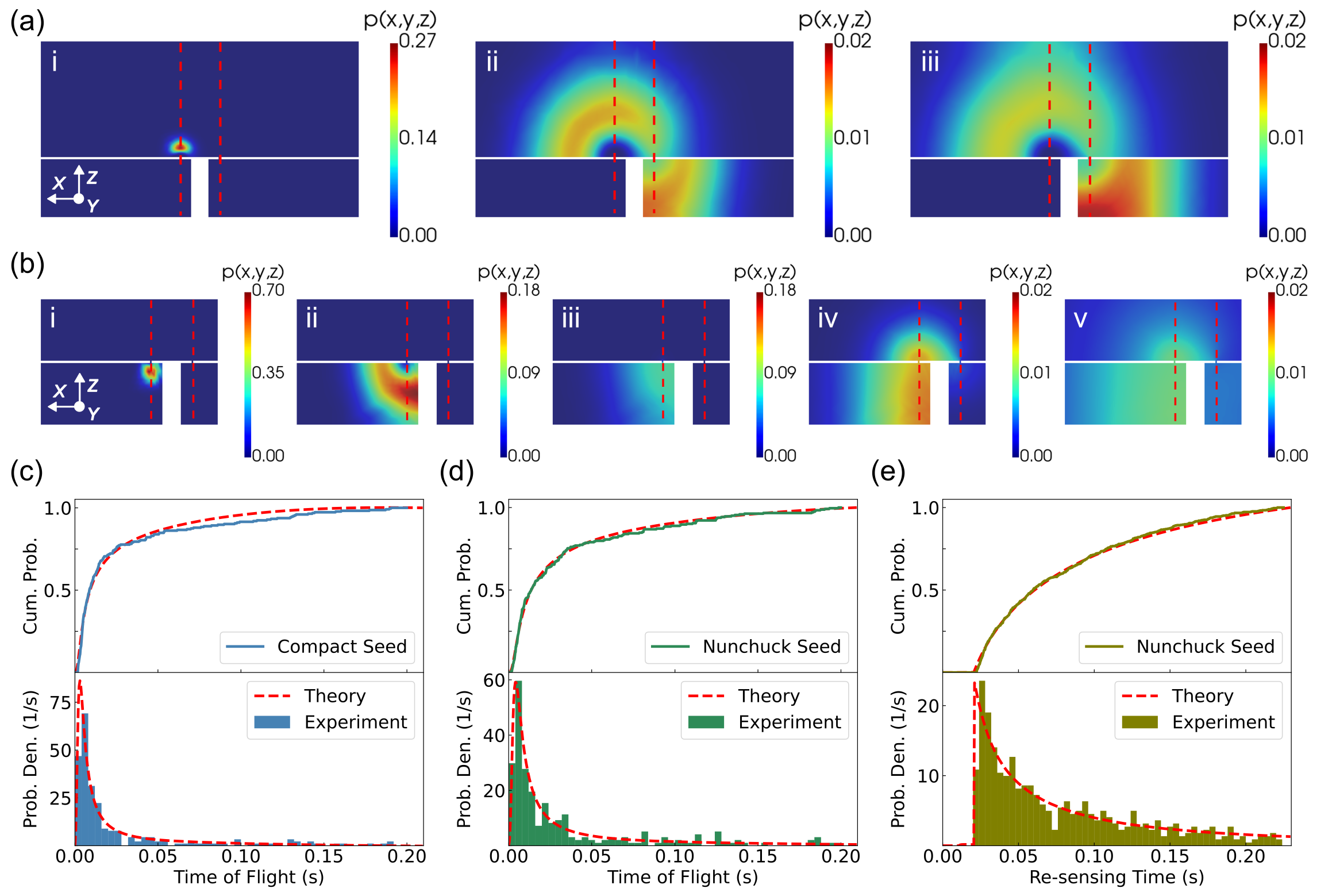} 
  \caption{Results from the finite element diffusion model for TOF and resensing process. 
  (a) Simulated PDF for the TOF process, displayed as color shading on a 2D vertical cross-section through the dual-pore arrangement.
  The $x$-axis is aligned along the pore separation axis and the $z$-axis is aligned perpendicular to the membrane surface.  
  Particles, initially distributed directly above pore 1 (i), migrate into the common chamber under the influence of diffusion and the electric field; those that approach pore 2 enter channel 2 (ii). 
  Over time, the number of particles in the common chamber diminishes and the number in channel 2 grows (iii).
  The dashed red lines highlight the nanopore locations.
  (b) Simulated PDF for the resensing process, displayed as color shading on a 2D vertical cross-section through the dual-pore arrangement.
  Particles, initially distributed directly below pore 1 (i), migrate further into channel 1 (ii). 
  After diffusing during the zero-voltage waiting time (iii), the particles are driven back into the common chamber via field reversal (iv) and undergo the TOF process towards channel 2 (v).
  Again, the dashed red lines highlight the nanopore locations. 
  (c-d, top) Cumulative distributions of modeled TOF (dashed red lines) that best fit the cumulative distribution of measured TOF for compact seed monomers (c, solid blue line)  and nunchuck seeds (d, solid green line).  
  (c-d, bottom) The corresponding probability density of modeled TOF (\textit{i.e.}, determined using parameter values obtained from fitting to the cumulative distribution, dashed red lines), compared to experimental TOF results for compact seed monomers (c, blue bars) and nunchuck seeds (d, green bars). 
  (e, top) Cumulative distribution of modeled resensing times (dashed red line) compared to the cumulative distribution of measured resensing times for nunchuck seeds.  
  (e, bottom) The corresponding probability density of modeled resensing time compared to the histogram of measured resensing times for nunchuck seeds.
}
  \label{model}
\end{figure}

The TOF distribution is the probability density (per unit time) of a particle that was re-sensed at pore 1 leaving the common chamber through pore 2. 
This can be evaluated from the rate of increase in the probability of finding the nanostructure in channel 2 (\textit{i.e.} PDF integrated over channel 2) at each time step. 
Figure~\ref{model}a gives side-view snapshots of the simulation.  
Under a constant electric field, the nanostructure gradually diffuses away from pore 1 (Fig.~\ref{model}(a-i)) and enters channel 2 through pore 2 (Fig.~\ref{model}(a-ii\&iii)). 

The modeled TOF distribution is then fit to the experimental distribution to obtain estimates of the diffusion constant $D$ and mobility $\mu$.  
To remove ambiguity regarding the choice of histogram binning, we fit the cumulative distributions.
Figure~\ref{model}(b,c) shows that the best fitting model closely follows the experimental TOF distributions for compact seed monomers and nunchuck seeds (results for looped seeds and fringed-looped cylinders are shown in Fig.~S8).  
  
This good agreement suggests that the particle transport dynamics during the TOF process is well represented by our model.  
Bootstrap resampling is used to assign an uncertainty of 10-20\% to the corresponding fitted diffusion constants (Table \ref{TOF}, Section S4.2). 
Our estimates of the electrophoretic mobilities, which intrinsically demand much more data to be precisely estimated, have much greater uncertainty: bootstrap resampling yields large variability ($\sim50$\%, Fig. S10).

In order to interpret the diffusion constants obtained from fitting, we estimate the corresponding seed length using the equation for the diffusion coefficient of a rod-like particle:
\begin{equation}
  D_{\text{rod}} = \frac{k_B T \ln(L/d)}{3 \pi \eta L} \label{rod}
\end{equation}
Here $\eta=0.89$ mPa·s is the solution viscosity\cite{viscosity}, $T=298$ K is the temperature and $d=12$ nm is the seed diameter.

We model our nanostructures as rigid rods because both the $\sim$70\,nm seed and the 11\,nm dsDNA linker in the nunchuck seed dimer are much shorter than their persistence lengths ($\sim$25\,$\mu$m~\cite{pl} and $\sim$50\,nm, respectively) and much longer than their diameters ($\sim$10\,nm and $\sim$2\,nm, respectively). 
For a more detailed justification of the rigid rod model, see Section S4.3.

In general, the estimated lengths are in good agreement with expectations based on nanostructure design, with deviations \textless15\%, consistent with the reported error-bars (Table 1). For the theoretical length estimate of the nunchuck, we provide a range that corresponds to including, and neglecting, the length of the 32\,bp (11\,nm) dsDNA linker. 
The linker, with a diameter of 2\,nm, is sufficiently short and thin compared to the compact seeds it connects (which are 12~nm in diameter) that it will contribute less overall friction than an equivalently long section of seed. Thus, modeling the nunchuck as a rigid rod with respect to its diffusion, we expect to extract an effective length that lies between the length estimates including or neglecting the linker length.

\begin{table}
  \caption{Diffusion coefficient and nanostructure length estimated from TOF distributions}
  \label{TOF}
  \begin{tabular}{lccc}
    \hline
    Nanostructure  & $D$\textsuperscript{\emph{*}} ($\mu\text{m}^2/\text{s}$) & estimated $L$\textsuperscript{\emph{*}} (nm) & theoretical $L$ (nm)   \\
    \hline
    Nunchuck   & 8.6$\pm$0.6 & 140$\pm$18 & 152-163   \\
    Compact & 12.7$\pm$1.1 & 66$\pm$14 & 76  \\
    Looped  & 11.8$\pm$1.6 & 78$\pm$22 & 76  \\
    Fringed-looped & 12.4$\pm$1.1 & 70$\pm$14 & 65 \\
    \hline
  \end{tabular}
  
  \textsuperscript{\emph{*}}  Errors represent standard deviations found by bootstrap resampling.
\end{table}

We apply a similar approach to analyze pore 1 resensing (Fig. \ref{model}(b)).  
In the resensing process, the nanostructure translocates through pore 1 and is then drawn back \textit{via} field reversal to translocate through pore 1 in the reverse direction.  
The pore 1 resensing process initiates immediately after the first translocation through pore 1 completes; here the nanostructure PDF at time $t=0$ is a Gaussian function centered at the bottom edge of pore 1 (Fig.~\ref{model}(b-i)). 
The PDF is then evolved \textit{via} Eq.~\ref{Smoluchowski} under electric fields specific to the resensing process. 
For the first 11\,ms of the process, the voltages are set to 300\,mV on channel 1 and -300\,mV on channel 2.
This step drives the nanostructure deeper into the channel (Fig.~\ref{model}(b-ii)), preventing immediate re-translocation after field reversal.
A 10\,ms waiting period ensues, during which all voltages are zeroed (Fig.~\ref{model}(b-iii)). 
This step, which also helps prevent translocation while current is unstable during the capacitive transient, allows diffusion to have a dominating influence on the time between sensing and resensing. 
Finally, the voltages are reversed to -300\,mV for channel 1 and 300\,mV for channel 2. 
Upon returning to the common chamber (Fig.~\ref{model}(b-iv)), the nanostructure initiates the TOF process directed towards pore 2 (Fig.~\ref{model}(b-v)).
The rate of increase of probability of nanostructures re-entering the common chamber at each timestep was taken to represent the probability distribution of resensing times. 

As for the TOF, we fit the simulation results to the measured distribution of resensing time delays.  
Figure \ref{model}(e) compares the best-fitting model resensing time distribution to the measured distribution for nunchuck seeds (equivalent results for compact seed monomers, looped seeds and fringed-looped cylinders are shown in Fig.~S9). 
The best-fitting diffusion coefficient and resulting seed length, as computed from Eq.~\ref{rod}, are shown in Table \ref{RTD}. 
Evidently, fitting to the distribution of resensing time delays yields large deviations (30-40\%) from the theoretical values.  
Likely, these higher deviations arise because of physical effects and artifacts introduced by the intrinsic need for time-varying fields in the resensing process. 
Dynamic reversal of field polarity is needed during resensing to ensure the constructs reverse direction and re-enter the pore. For example, in the model, it is assumed that nanostructures returning to the common chamber will be observed as soon as the reversed voltages are applied. 
However, capacitive transients impose a finite current settling time of $\sim$~5\,ms following voltage reversal, during which translocation cannot be detected.  
Simulation suggests that about 9\% of events  occur within this 5\,ms time window.
The absence of these early time points is likely responsible for some of the discrepancy between the theoretical nanostructure lengths and those estimated from resensing time distributions. The capacitive effects also lead to gradually evolving electric fields upon voltage reversal, while the model assumes instantaneous field changes. 
Although refinement of the resensing time delay model may improve agreement, the consistently strong results obtained with the simpler TOF process highlight the advantages of the dual-nanopore approach.
\begin{table}
  \caption{Diffusion coefficient and nanostructure length estimated from resensing time distributions}
  \label{RTD}
  \begin{tabular}{lccc}
    \hline
    Nanostructure  & $D$\textsuperscript{\emph{*}} ($\mu\text{m}^2/\text{s}$) & estimated $L$\textsuperscript{\emph{*}} (nm) & theoretical $L$ (nm)   \\
    \hline
    Nunchuck   & 6.3$\pm$0.5 & 230$\pm$30 & 152-163   \\
    Compact & 13.7$\pm$1.2 & 54$\pm$11 & 76  \\
    Looped  & 13.4$\pm$1.8 & 57$\pm$16 & 76  \\
    Fringed-looped & 14.0$\pm$1.2 & 50$\pm$10 & 65 \\
    \hline
  \end{tabular}
    
  \textsuperscript{\emph{*}}  Errors were estimated using the same scale as in the TOF experiment.
\end{table}

\subsection{Multi-translocation trajectory of single seeds}

In addition to the triple-scan events used in the machine learning classifications, the nanostructure has the possibility of translocating back to the common chamber through pore 2 (Fig.~2(v)) and undergoing another resensing cycle. 
A single nanostructure can thus be sensed more than three times, producing higher order multi-translocation events.  We display such higher order events in the form of a trajectory plot, where each translocation event associated with a single structure is plotted as a point in (conductance blockade, dwell time) space,  and connected to the point corresponding to the subsequent event for that structure.

\begin{figure}[h!]
    \centering
    \includegraphics[width=1\linewidth]{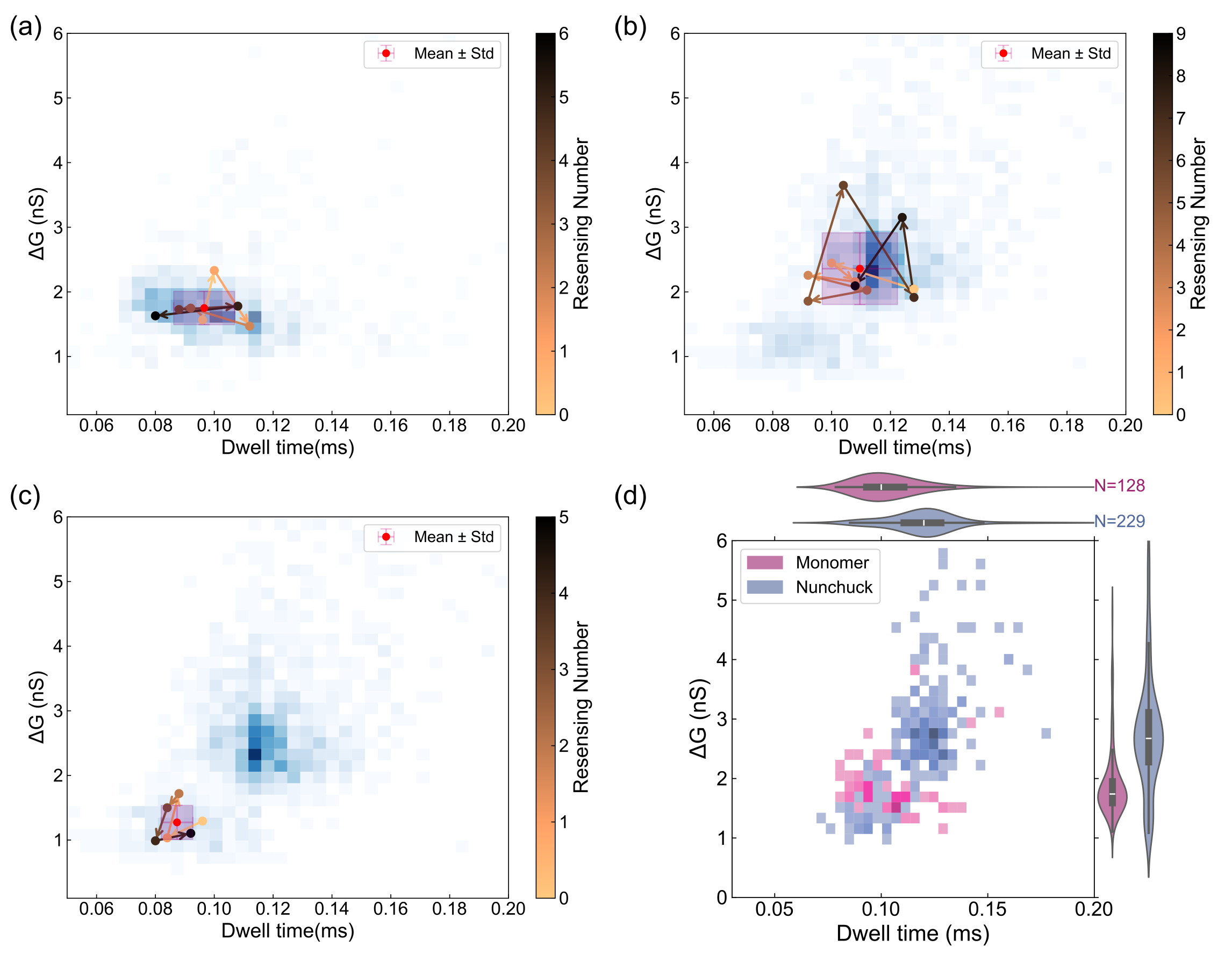} 
  \caption{Multi-translocation trajectory of a compact seed monomer (a), a nunchuck seed (b), and a compact seed monomer in the nunchuck sample (c), with background heatmap indicating the distribution of all translocation events for the corresponding samples.  
  Trajectories highlight the variation in the conductance blockade and dwell time for the same molecule during multiple translocations (order of translocation events is indicated by point color, with lighter color indicating earlier translocations). 
  The red point marks the mean of all events in the trajectory; the error bars give standard deviations. 
  (d) Density heatmap of mean conductance blockade and dwell time over multi-translocation events for each single structure for compact and nunchuck sample, with corresponding marginal violin plots. Only constructs achieving RN 3 or higher are included here.
 }
  \label{trajectory}
\end{figure}

Figure~\ref{trajectory}(a,b) show examples of the (conductance blockade, dwell time) trajectories of a compact seed monomer and a nunchuck seed, respectively. 
The trajectory starts at the first sensing event for pore 1 (resensing Number (RN) 0).  
The next point is the resensing at pore 1  (RN 1), followed by pore 2 sensing (RN 2), with subsequent events indicated by gradually deepening color (\textit{e.g.}, RN 4 refers to the event when the molecule is captured by pore 1 again, starting a second cycle).  
The highest RN we achieved for this structure is RN 9 (\textit{i.e.}, the molecule did not escape until it had traveled from pore 1 to pore 2 for the third time).  
The background density heatmap indicates the distribution of all translocation events (through either pore) for the corresponding sample. 
The diameters of pore 1 and pore 2 are sufficiently similar that events from the two pores have similar distributions (Fig.~S12). 

In agreement with their corresponding background distributions, compact seed monomer trajectories are more localized and nunchuck seed trajectories are more spread out.  
This makes qualitative sense because nunchuck seeds consist of two compact seed monomers linked end-to-end \textit{via} a 32\,bp dsDNA linker.  
While fluctuations in compact seed monomer translocation behavior arise mainly from translational Brownian motion and different orientations assumed upon entering the pore, nunchuck seeds experience additional translocation modes arising from linker bending and twisting.
 
As mentioned previously, two clusters are observed in the heatmap of the nunchuck sample, and we attribute the lower left cluster to the presence of compact seed monomers.  
We therefore expect a subset of the trajectory plots, corresponding to monomers, to overlap exclusively with the lower left cluster.  
We do, indeed, observe such events (Fig.~\ref{trajectory}c), demonstrating that trajectory plots can help detect impurities or profile secondary populations of structures in a sample at the level of individual analytes.
 
Averaging over the multiple translocation events of a single structure helps minimize the effect of fluctuations and can aide in discriminating chemical and structural differences between analytes. 
This is seen in the distribution of mean conductance blockades and dwell times of each single nanostructure (Fig.~\ref{trajectory}(d), Fig.~S13). 
After averaging, the distributions are more localized than in the single-pore heatmap in Figure~\ref{violin}. 
 
\section{Conclusion}

Dual nanopore chips operated with active logic enable multiple translocations of the same nanostructure through and between the pores, increasing the number and type of measurements that can be performed on a given nanostructure (\textit{i.e.} enabling multiple measurements of conductance blockade/dwell time as well as access to TOF and resensing time).  Here we use this capability, combined with machine learning-based classification, to discriminate between four structurally similar DNA origami with accuracies of greater than 80\%.  For example, we are able to discriminate between structures with a 12\,nm length difference and/or distinguish 336-base ssDNA loop from its folded six-helix bundle counterpart. These structures are too similar to identify via single-pore measurements with pores of the size used. In addition, nanostructure time-of-flight (TOF) between the pores, combined with a model that incorporates diffusional dynamics, can yield quantification of a nanostructure's diffusion constant and length. 
This approach enhances the capability of nanopore sensing and its utility for the field of DNA nanotechnology.

This study intentionally focuses on a set of structurally similar nanostructures to highlight the discrimination capability of our dual-pore system.  Given the wide range of DNA nanostructures already characterized \textit{via} single solid-state pores\cite{DNAstar, cube-ring, four-brick, wang2019current, He2023, zhu2018adaption, zhu2020deformation}, we argue that our technique is potentially broadly applicable to other DNA nanostructure classes.  In principle, nanostructures successfully characterized using a single-pore can be characterized at equivalent bandwidth using a dual-pore device possessing a second pore of comparable dimensions. Use of our active dual-pore setup will then \emph{a priori} yield a higher dimensional parameter space for these structures that will increase discrimination accuracy. The key technical limitation of our current approach is that the nanopore size is limited by the FIB drilling protocol used to around 25\,nm.  Thus, smaller nanostructures detectable with $\sim$10\,nm single pores may not be resolved by our \textgreater25\,nm pores.  Smaller pores may be fabricated \textit{via} other techniques in the future, such as tip‐controlled local breakdown (TCLB)\cite{TCLB}, to address this limitation.

Compared to conventional techniques already widely employed to analyze DNA nanostructures (\textit{e.g.} gel electrophoresis, AFM and CEM), our dual-pore platform can output quantitative single structure level information with increased operational accessibility.  Unlike CEM and AFM, which are usually housed in specialized facilities due to their expensive instrumentation and need for expert-level operation, our system is accessible to smaller labs due to its lower cost, smaller foot-print and relative ease-of-use.  While gel electrophoresis systems possess comparable accessibility, gel electrophoresis provides only bulk (ensemble-averaged) information. In contrast, our dual-pore approach, \textit{via} enabling repeated measurements over single molecules, provides quantitative single structure level estimates of nanostructure size.   Moreover, the dual-pore technique requires minimal sample preparation and enables real-time sensing under bulk solution.

A key challenge that can be addressed via further refinement of the technique is that, during the resensing process, the nanostructure has a probability of diffusing away (\textit{i.e.}, escaping) from the access region of the nanopores. 
This leads to difficulty in obtaining trajectories with high resensing numbers. 
We observe a lower probability of escape for nunchuck seeds than for compact seed monomers (Fig.~S14) due to the lower diffusion coefficient of nunchuck seeds. 
We hypothesize that the escape probability may be lowered by reducing the pore-to-pore distance.  
An alternative strategy might be to add a `guiding field' in the common chamber, either hydrodynamic or electrical in origin, that would impose a convective bias on the inter-pore transport, reducing diffusional escape while retaining sensitivity to diffusional dynamics.  
If realized, this strategy may yield multi-translocation events with higher RN that, in addition to increasing classification accuracy, would potentially enable extraction of diffusion constants for individual structures.  

In the future, we suggest that this dual-pore technique may enable tracking complex nanostructure size distributions in the solution environment.  For example, during dynamic self-assembly of DNA nanostructures, such as tile-based DNA nanotubes, samples can contain structures with a wide-variety of different sizes/morphologies arising from distinct assembly pathways\cite{seed}.  To analyze these intrinsically heterogeneous samples, there is a need to reduce measurement noise arising from intrinsic heterogeneity in the translocation process.  The dual-pore approach's capability of increasing single structure measurement statistics is a natural way to overcome this challenge.  Note that the measurement time-scale is sufficiently fast that obtaining increased single structure statistics does not sacrifice the ability to profile large numbers of distinct structures on a practical time-scale.  For example, we find a measurement time-scale of $\sim 0.2$ sec per each additional pore traversal/resensing, suggesting that roughly five additional resensing measurements can be performed for each additional second of capture (so roughly 10 re-samples over 2\,s).  Practically, concentrations must be kept sufficiently low to avoid engaging more than one nanostructure in the dual-pore arrangement at a time.  Specifically, this implies a time-scale for capturing distinct structures about 10$\times$ larger than the time-scale for analyzing a given single structure.  With these constraints, it is possible to sample about three separate nanostructures per minute with around 10 translocations each, leading to around 100 nanostructures analyzed for half an hour operation.  Furthermore, we note that, by systematically varying the relative diameter of the two pores, it may be possible to explore how pore geometry alters translocation dynamics and further increase the information that can be extracted from a single nanostructure.

\section{Methods}\
\subsection{Nanopore fabrication}
The nanopore chips used in this study were fabricated following the procedure described in Ref.~\citenum{small2018}. 
Briefly, two wafers were used: one glass and one silicon (Si). 
Two flat-bottomed, "V"-shaped microchannels were dry-etched into the glass wafer, such that their tips were separated by 0.4\,$\mu$m (Fig.\,\ref{chip}). 
A multilayer stack was deposited on the Si wafer, consisting of a layer of 400\,nm of SiN (deposited by low-pressure chemical vapor deposition, LPCVD), a layer of 100\,nm of \ce{SiO2} (deposited by plasma-enhanced chemical vapor deposition, PECVD) on one side (the frontside), and another layer of 30\,nm SiN (also by LPCVD). 
Note that the Si wafer's backside was coated with 430\,nm of SiN from the LPCVD step.
The Si wafer was then anodic bonded to the glass wafer, with the \ce{SiO2}-containing stack facing the microchannels. 
The backside SiN deposited during the LPCVD process was removed via dry etching, and the Si substrate was subsequently removed using hot KOH solution. 
Then, to expose the SiN membrane, a 10\,$\mu$m$\times$10\,$\mu$m window was defined by dry etching through the 400\,nm SiN layer and partially into the oxide, followed by selective wet etching of the remaining \ce{SiO2}. 
Finally, two nanopores were drilled through the SiN membrane at the microchannel tips using focused ion beam (FIB) milling.

\subsection{DNA origami assembly}
DNA origami were folded by thermally annealing a 1:10 mixture of scaffold:staple strands in 1xTAE$\cdot$Mg buffer (40\,mM Tris, 12.5\,mM magnesium acetate, 1\,mM ethylenediaminetetraacetic acid (EDTA), pH 8.3). 
DNA strand sequences are listed in Supplementary Information. 
The p3024 scaffold was purified in house from XL1-Blue MRF' supercompetent cells (cat. no. ST200230, VWR) doubly transformed with pScaf-3024.1 scaffold plasmid\cite{Douglas} and HP17\_KO7 helper plasmid\cite{Praetorius2017} (Addgene plasmid\,\#s 111404 and 120346, respectively). 
Staple strands were purchased with standard desalting from Integrated DNA technologies, Inc. (Coralville, IA, USA). 
The nunchuck seed linker strands were kinased before annealing, then ligated after annealing using the same protocol as in a previous study~\citenum{cai2020}.
All nanostructures were purified by electrophoresis through a 1\% agarose gel containing 0.5\,$\mu$g/mL ethidium bromide in 1xTAE$\cdot$Mg buffer. 
The nanostructure bands were cut out of the gel, frozen at -20 °C for 5 min, placed in Freeze 'N Squeeze\textsuperscript{TM} DNA gel extraction spin columns (cat. no. 7326166, Bio-rad) and spun for 3 minutes at $13,000\times g$ at room temperature. 
Purified nanostructures were stored at room temperature for up to three weeks before being diluted and loaded into the nanopore apparatus.

\subsection{Nanopore measurements}
To run the experiments, the dual-nanopore chip was mounted to a custom-designed chuck fabricated by Protolabs (see Supplementary Information, Section 8). 
A homemade PDMS gasket (0.8\,mm thick) between the chuck and the surface of the chip ensured leakproof sealing. 
Measurements were performed in aqueous 1\,M LiCl with 10\,mM Tris and 1\,mM EDTA at pH 7.9. 
A current amplifier (Multi-Clamp 700B, Molecular Devices) was used to collect both current and voltage signals, and an analog-to-digital converter (Axon Digidata 1550B, Molecular Devices, San Jose, CA) was used to digitize the data. 
The active feedback control was built on an FPGA (National Instruments PCIe-7851R, Emerson, Austin, TX), and its algorithm was developed using LabView (National Instruments, v20.0f1).

\begin{acknowledgement}

W.R acknowledges support from the Natural Science and Engineering Research Council of Canada (NSERC RGPIN-2018-06125), the Fonds de recherche du Quebec-Nature et technologies (FRQNT-NT, PR-286442) and NIH award number 1R21HG011236-01.  
D.K.F acknowledges support from the U.S. National Science Foundation (FMRG:Bio award number 2134772). 
R.L. was supported in part by an NSF Research Traineeship (NRT award \#2152201). 

\end{acknowledgement}

\begin{suppinfo}
Gel electrophoresis results and additional AFM of the nanostructures; oxDNA simulation parameters; confusion matrices from machine learning classifications; definition of machine learning performance metrics; details on the training of the classifier; TOF and resensing time fitting results; details of the finite element model; suitability of the isotropic, rigid rod diffusion model; density heatmap for translocation events through pore 1 and pore 2; heatmap comparison for fringed-looped cylinders, looped seeds and compact seed monomers after averaging; number of translocation events for each resensing number; recipe and protocols for DNA nanostructure self-assembly; sequences of the DNA strands; chuck design; SEM image of the dual-nanopore chip.

\end{suppinfo}

\bibliography{references}

\end{document}


\begin{center}
    {\Large {\bf Supplemental Information} for: \\{\bf DNA nanostructures characterized via dual nanopore devices}}
\end{center}

\begin{center}
{\large Wangwei Dong$^{1,*}$, Zezhou Liu$^{1,*}$, Ruiyao Liu$^{2}$, Deborah Kuchnir Fygenson$^{2,3}$, Walter Reisner$^{1}$\\[1em]}
{\normalsize $^{1}$Department of Physics, McGill University, Montr\'eal, Qu\'ebec, Canada, H3A 2T8\\
$^{2}$Department of Physics, University of California, Santa Barbara, CA 93106\\
$^{3}$interdisciplinary program in Quantitative Biology, University of California, Santa Barbara, CA 93106\\
$^*$ These authors contributed equally to this work.\\[1em]
E-mail: fygenson@ucsb.edu; walter.reisner@mcgill.ca}
\end{center}
\newpage

\section{Gel electrophoresis and AFM}

\subsection{Nanostructure purification by agarose gel electrophoresis}
\begin{figure}[h!]
    \centering
    \includegraphics[width=0.5\linewidth]{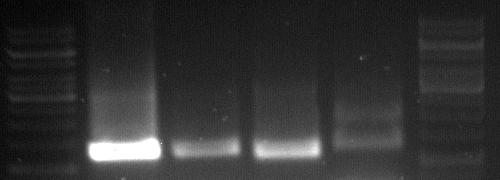}
    \caption{Annealed nanostructures were run on an 1\% agarose gel with EtBr in 1$\times$TAE$\cdot$Mg at 110~V for 90 min. The first and last lanes contain 1~kb ladder DNA (Gene ruler, ThermoFisher). The lanes in between, from left to right, correspond to compact seed monomers, looped seeds, fringed-looped cylinders, and nunchuck seeds. The first three nanostructures were purified by cutting out the bright bands. Nunchucks seeds were purified by cutting out the upper band.}
    \label{fig:gel-EtBr}
\end{figure}
    
\begin{figure}[h!]
    \centering
    \includegraphics[width=0.5\linewidth]{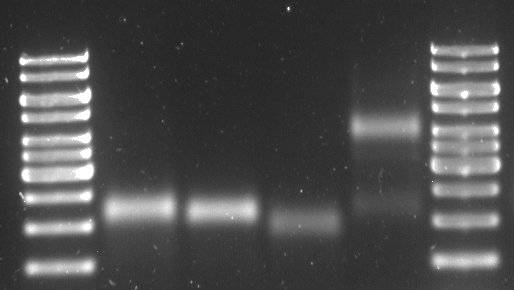}
    \caption{Gel purified nanostructures used in the experiments. Lane descriptions are the same as in Fig. S1: between the ladders, from left to right, are compact seed monomers, looped seeds, fringed-looped cylinders, and nunchuck seeds. This 1\% agarose gel was run at 120\,V for 90\,min in 1$\times$TAE buffer (without Mg) and post-stained with SYBR gold (cat no. S11494, Thermo Fisher).}
    \label{fig:gel-assay}
\end{figure}
\newpage
\subsection{Nunchuck seed samples contain compact seed monomers.}
\begin{figure}[h!]
    \centering
    \includegraphics[width=0.7\linewidth]{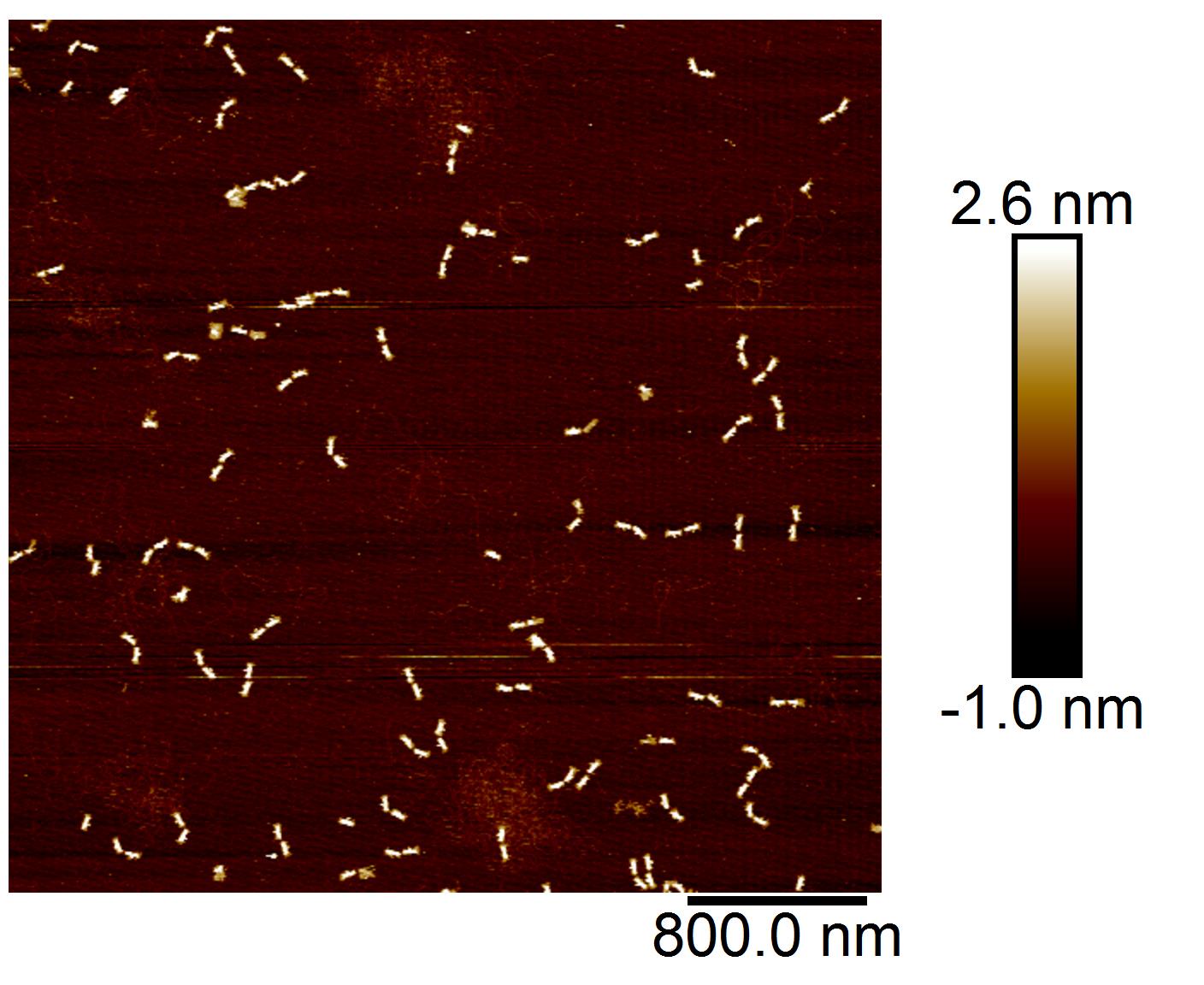} 
  \caption{AFM image of the nunchuck seed sample. Note that $\thicksim12$\% of the nanostructures correspond to compact seed monomers.
 }
  \label{AFM}
\end{figure}

\section{oxDNA simulation parameters}

Nanostructures were encoded using scaDNAno\cite{scadnano} and relaxed in two stages,\cite{Poppleton2023} using DNA interaction type specified in P. Šulc \emph{et al}\cite{Sulc2012Sequence} with parameters max\_backbone\_force = 5 and verlet\_skin = 0.15. 
The first relaxation stage minimized the local potential energy with sim\_type=min, which involved 2000 steps of size $dt = 0.005$ at 300\,K. 
The second relaxation stage was also performed at 300\,K, and took $10^5$ steps of size $dt =  0.0075$ with MD simulation mode set to Langevin thermostat (sim\_type=MD) and diff\_coeff = 2.5.

\newpage
\section{Machine learning classification of dual pore translocations}
\subsection{Confusion matrices}
\begin{figure}[h!]
    \centering
    \includegraphics[width=0.4\linewidth]{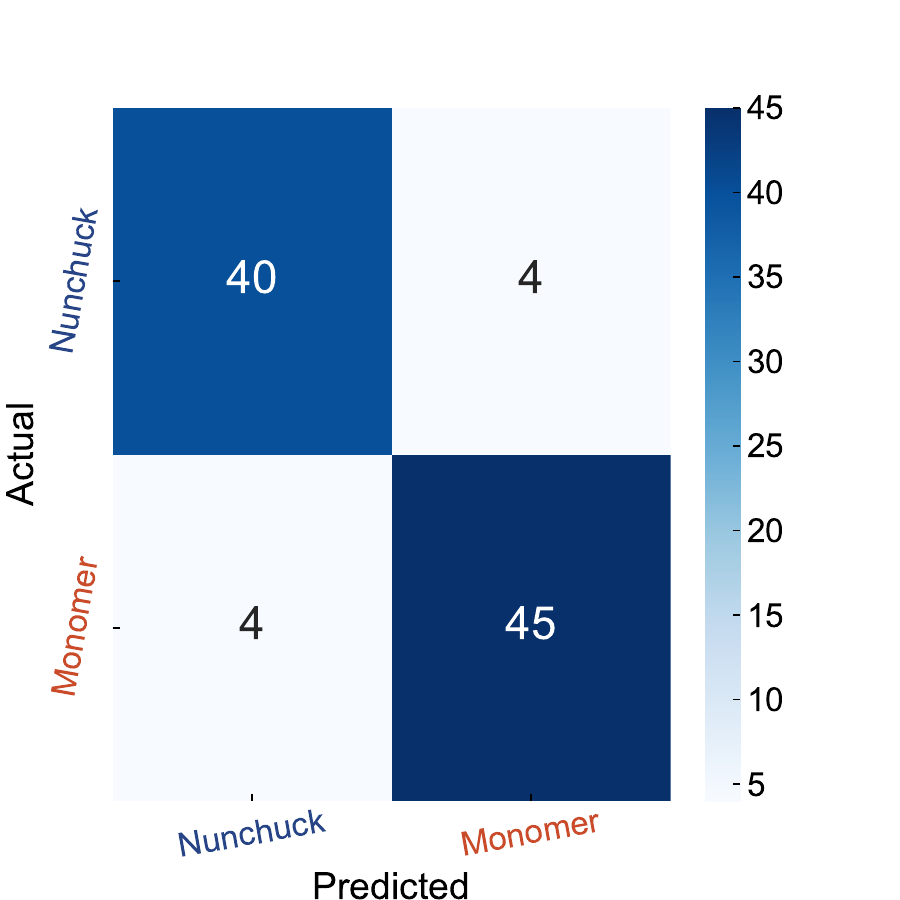} 
  \caption{Confusion matrix for identification of nunchuck seeds and compact seed monomers using only results from single translocations through pore 1 (pore 1 sensing events).
 }
  \label{nc}
\end{figure}

\begin{figure}[h!]
    \centering
    \includegraphics[width=0.8\linewidth]{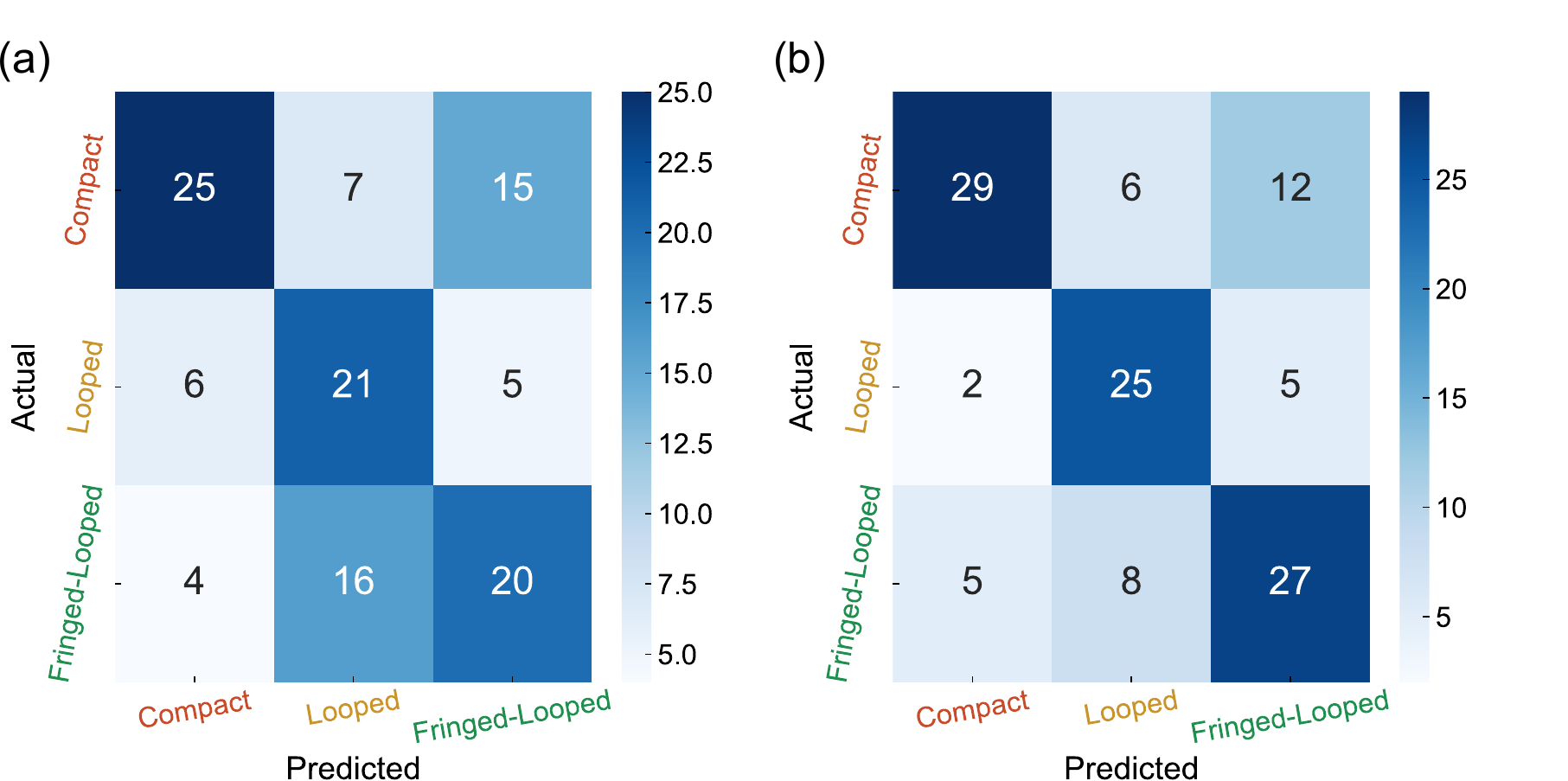} 
  \caption{Confusion matrix for identification of compact, looped, and fringed-looped seeds using results from (a) single-scan (pore 1 sensing events) and (b) double-scan (pore 1 sensing and resensing events), with a 30\,kHz bandwidth.
 }
  \label{10khz}
\end{figure}

\begin{figure}[h!]
    \centering
    \includegraphics[width=1\linewidth]{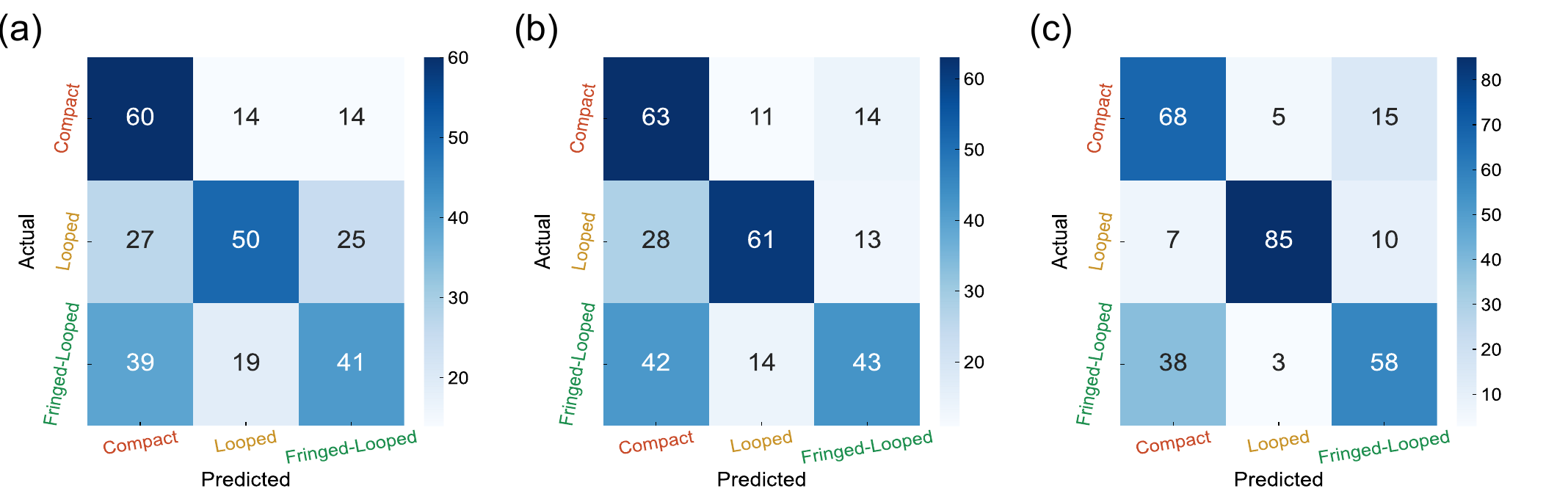} 
  \caption{Confusion matrix for identification of compact, looped, and fringed-looped seeds using results from (a) single-scan (pore 1 sensing events), (b) double-scan (pore 1 sensing and resensing events), and (c) triple-scan (pore 1 sensing, pore 1 resensing \& pore 2 sensing events), with a 10\,kHz bandwidth.
 }
  \label{10khz}
\end{figure}

\begin{figure}[h!]
    \centering
    \includegraphics[width=0.6\linewidth]{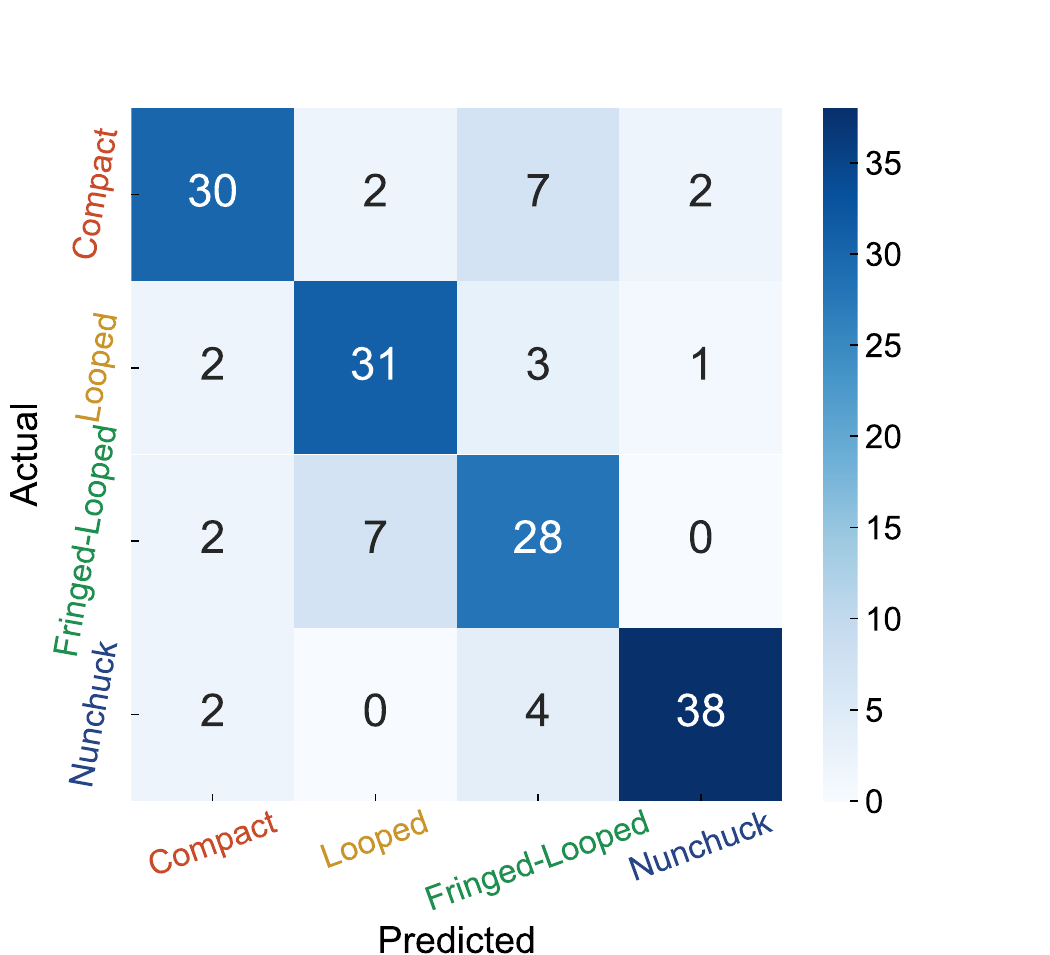} 
  \caption{Confusion matrix for identification of compact, looped, fringed-looped, and nunchuck seeds using triple-scan results with a 30\,kHz bandwidth.
 }
  \label{four-seed}
\end{figure}

\newpage
\subsection{Definition of machine learning performance metrics}
In random forest classification, three performance metrics were calculated for each sample individually: precision, recall and F1-score. 
Precision refers to the proportion of correctly predicted positive instances (True Positives, TP) out of all predicted positive instances ({\em i.e.}, including False Positives, FP), which is 
\[
\text{Precison}=\frac{TP}{TP+FP}.
\]
Recall refers to the proportion of correctly predicted positive instances (TP) out of {\em all} predicted positive instances ({\em i.e.}, including False Negatives, FN), which is 
\[
\text{Recall}=\frac{TP}{TP+FN}.
\]
F1-score balances precision and recall, and is defined as
\[
\text{F1-score}=2 \times \frac{\text{Precision}\times \text{Recall}}{\text{Precision} + \text{Recall}}.
\]
In this work, ``average F1-score'' refers to the mean of the F1-scores across all samples.

Accuracy is a metric used to evaluate the overall performance of the classification. 
It is defined as the proportion of correctly classified instances (TP and True Negatives, TN) among all instances in the dataset. That is,
\[
\text{Accuracy}=\frac{TP+TN}{TP+TN+FP+FN}.
\]
Because the datasets reported here were all balanced ({\em i.e.}, all classes had a similar number of events), accuracies and average F1-scores were always similar in value.

\subsection{Details on the training of the classifier}

Our classifier is based on the Random Forrest (RF) approach \cite{breiman2001random}.  The first step in constructing an RF classifier is to build an ensemble of random trees, with each tree a decision structure that predicts the class (in our case DNA nanostructure) corresponding to input data (measured parameters for a given three-scan event).  The final decision is then averaged over all trees.

The two key hyper-parameters of the random forest (RF) algorithm are (i) the number of estimators (number of trees) and (ii) the maximum depth of the estimators (the depth of each decision tree).  In the main text, we chose max\_depth=10 and n\_estimator=100 as a balance between under and overfitting.  In order to determine how strongly classifier performance depends on choice of hyper-parameter, we performed a grid scan of these two parameters (varying max\_depth from 4 to 15 and n\_estimator from 50 to 150). The accuracy was not very sensitive to either hyper parameter.
The average accuracy over the hyper parameter range was 0.80~$\pm$~0.03, where the uncertainty corresponds to one standard deviation.

Randomness is a fundamental aspect of the RF algorithm, used to reduce correlation among the trees. 
Randomness is introduced to the tree structure by 1) bootstrapping the training set (randomly sampling with replacement); and 2) randomly selecting a subset of features. The training data determines the tree structure via recursive selection of the feature and threshold that best split the data into the purest possible datasets at each node (the best split strategy), a process explicitly defined in the original work by Breiman (2001) \cite{breiman2001random}.
Implementation of the RF approach involves:
\begin{itemize}
\item defining the number of estimators and the max depth of the estimators; 
\item feeding the training data set into the classifier; 
\item inputting the testing data into the fitted RF classifier and obtaining the predicted class; and 
\item comparing the predicted class with the ground truth. 
\end{itemize}
We evaluate the performance of the classifier by accuracy, f1 score and the confusion matrix.

\newpage

\section{Estimating molecular length from diffusion dynamics modeling}
\subsection{TOF and re-sensing fitting results}
\begin{figure}[h!]
    \centering
    \includegraphics[width=1\linewidth]{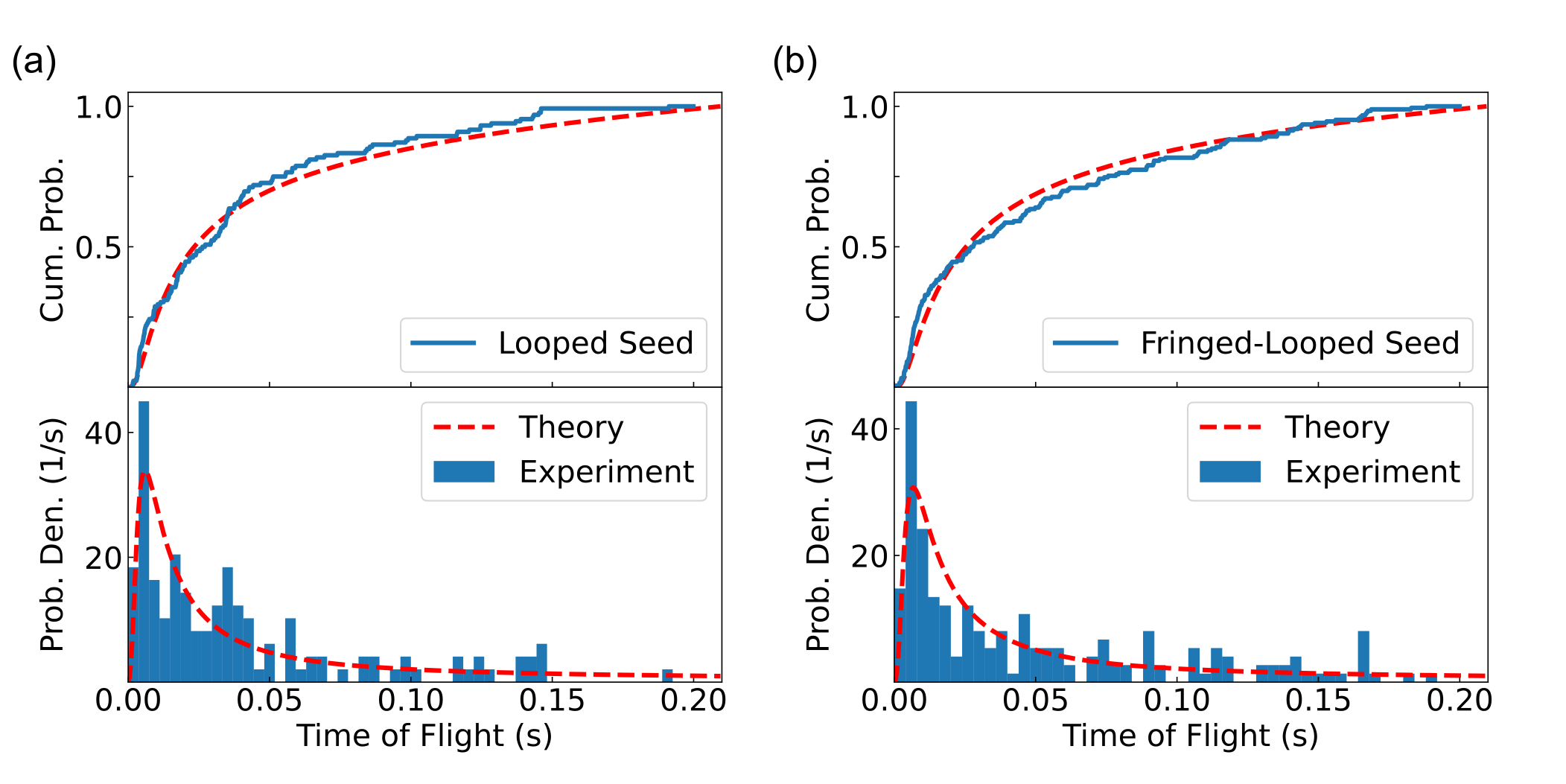} 
  \caption{Results from the finite element diffusion model for time-of-flight (TOF). (top) Best fitting cumulative distribution of modeled TOF (dashed, red) compared to the cumulative distribution of experimental TOF for (a, top) looped seeds and (b, top) fringed-looped cylinders.  (bottom) The corresponding modeled TOF distribution ({\em i.e.}, determined using parameter values obtained by fitting to the cumulative distribution), compared to experimental TOF results for (a, bottom) looped seeds and (b, bottom) fringed-looped cylinders.
 }
  \label{seed46}
\end{figure}

\begin{figure}[h!]
    \centering
    \includegraphics[width=1\linewidth]{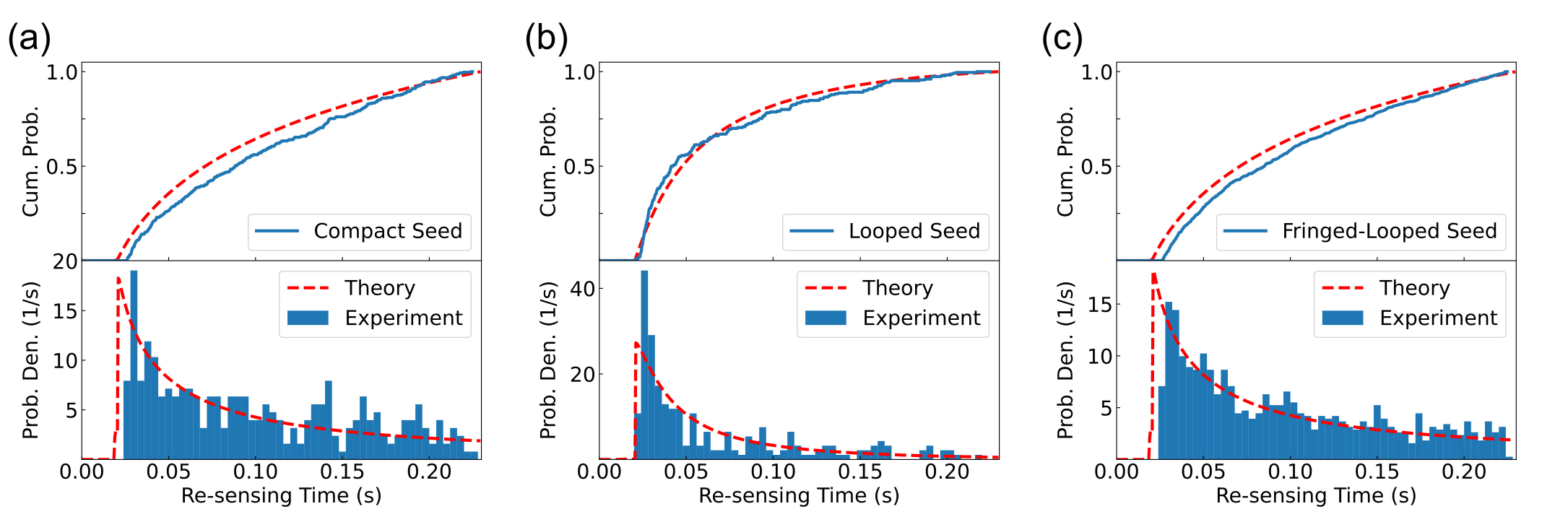} 
  \caption{Results from the finite element diffusion model for resensing.(top) Fits of the cumulative distribution of modeled resensing times to the cumulative distribution of measured resensing times for (a, top) compact seeds, (b, top) looped seeds and (c, top) fringed-looped cylinders.  (bottom) The corresponding modeled resensing time distribution ({\em i.e.}, determined using parameter values obtained from fitting to the cumulative distribution), compared to experimental resensing results for (a, bottom) compact seeds, (b, bottom) looped seeds and (c, bottom) fringed-looped structures.
 }
  \label{recapture-seed46}
\end{figure}

\begin{figure}[h!]
    \centering
    \includegraphics[width=0.5\linewidth]{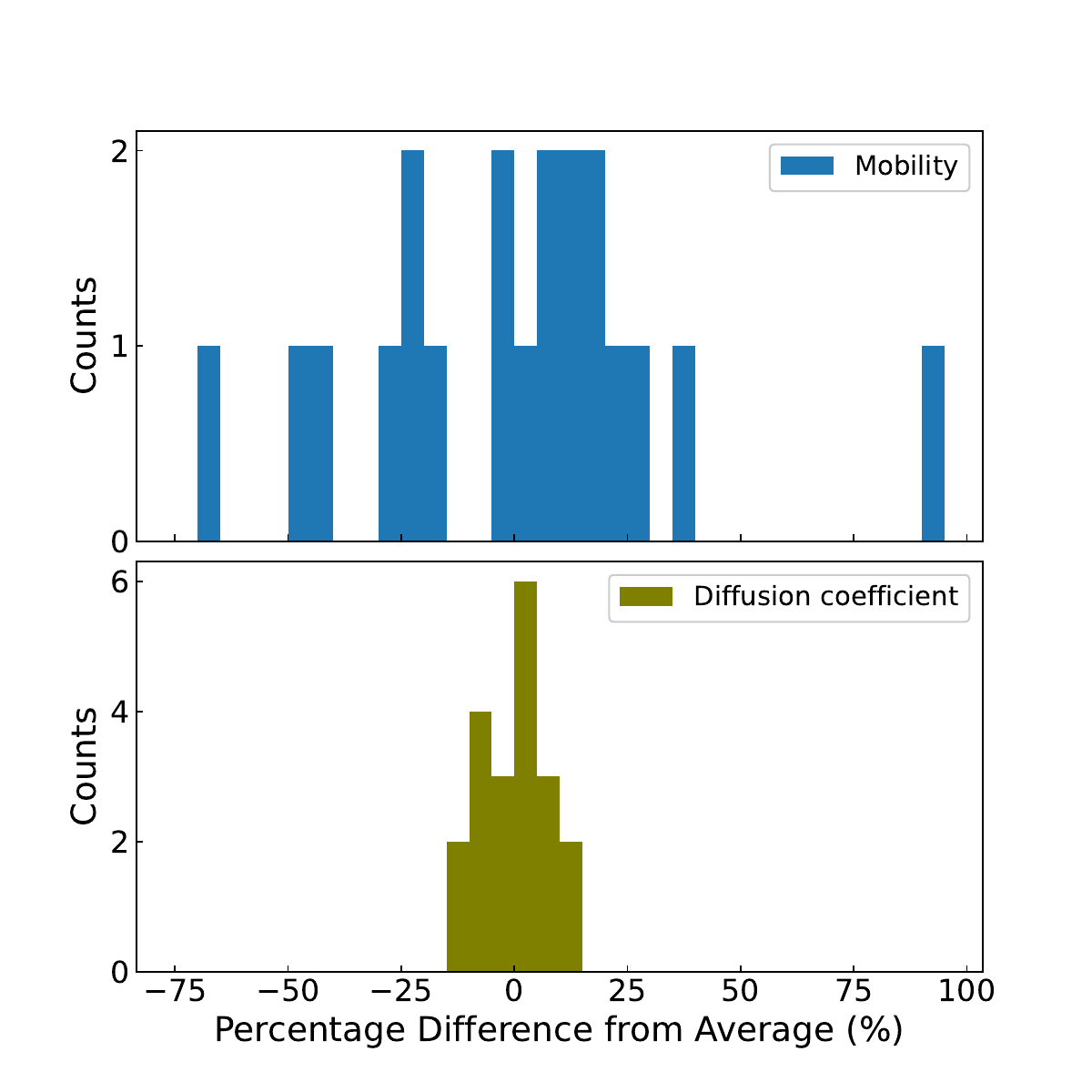} 
  \caption{Distributions of the fitted mobility (top) and diffusion coefficient (bottom) for nunchuck seeds over 20 synthetic datasets from bootstrap resampling, on a scale of percentage difference from average.
 }
  \label{mobility-distribution}
\end{figure}

\subsection{Details of the finite element diffusion model}
\begin{figure}[h!]
    \centering
    \includegraphics[width=1\linewidth]{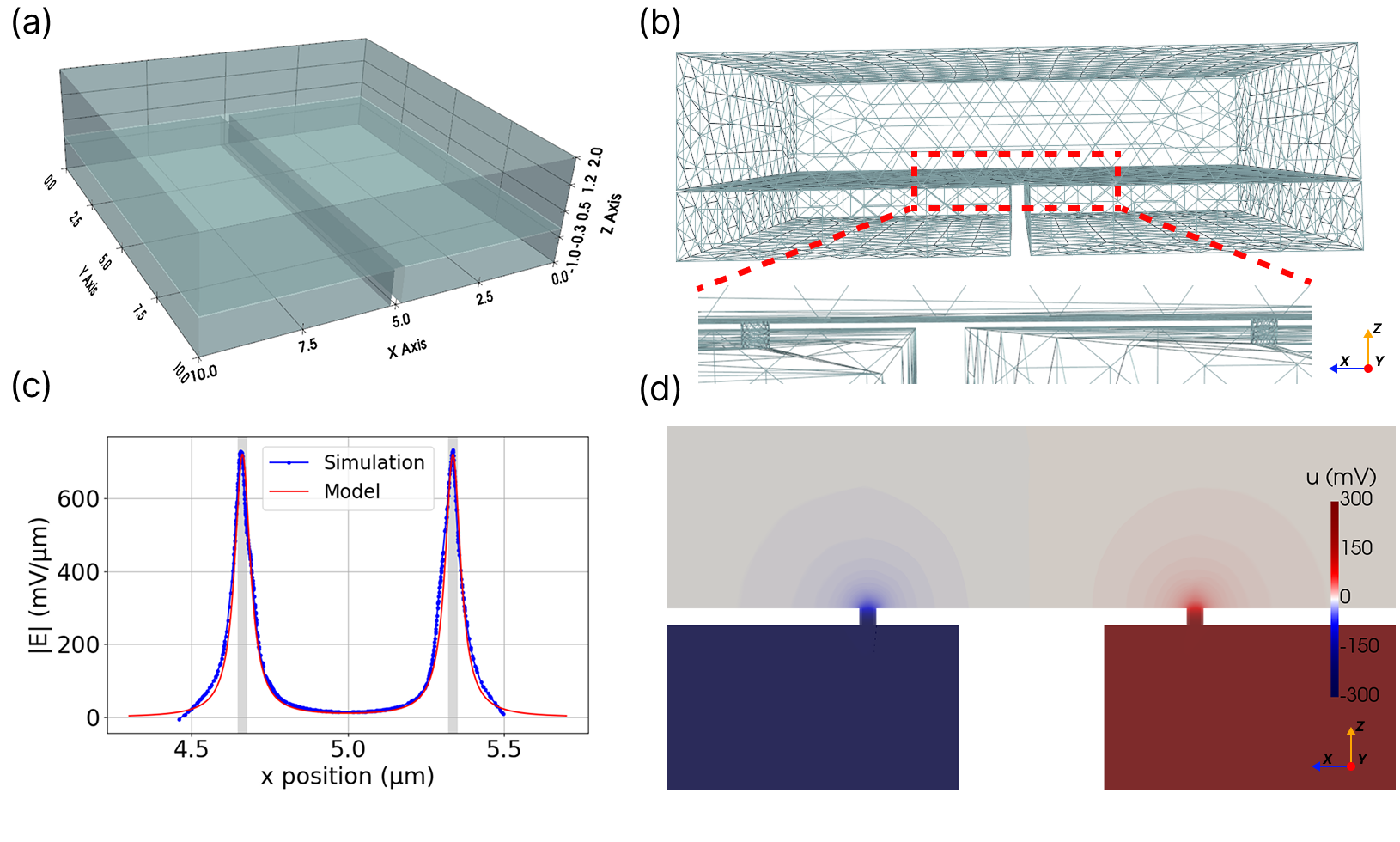} 
  \caption{(a) Model geometry in units of $\mu$m. (b) 3D mesh of the dual pore device, with zoomed-in image of the dual nanopore region. (c) Electric field along x axis at one pore diameter (30\,nm) above the surface (blue), compared to a dipole field model (red). The position of nanopore areas are shaded in grey. (d) Cross-sectional and zoomed-in view at y = 5 of the electric potential distribution in the dual nanopore region, when channel 1 is set at -300\,mV and channel 2 is set at +300\,mV.
 }
  \label{mesh}
\end{figure}

The geometry of the dual-nanopore device was modeled in FreeCAD and the corresponding mesh was generated by Gmsh, as shown in Figure \ref{mesh}(a, b).

The simulation was implemented using the finite element method (FEM) in FreeFem++. 
Particles were initialized with a three-dimensional, isotropic Gaussian probability density function (PDF) centered near the nanopore entrance:
\[
f(x, y, z) = \frac{1}{(\sqrt{2\pi}\cdot \sigma)^3} 
\exp\left( - \frac{(x - x_0)^2}{2\sigma^2} - \frac{(y - y_0)^2}{2\sigma^2} - \frac{(z - z_0)^2}{2\sigma^2} \right)
\]
Here, $\sigma=0.0008$ is the spread of the distribution and $(x_0,y_0,z_0)$ is its initial center position. 
For TOF modeling, we set $(x_0,y_0,z_0)=(5.337,5,0.001)$; for resensing modeling, we set $(x_0,y_0,z_0)=(5.337,5, -0.031)$.

The Poisson equation was solved assuming the glass and SiN surfaces are hard walls with Dirichlet and Neumann boundary conditions, respectively:
\[
  \phi_\Gamma = V; \;\;  \qquad \left.\frac{\partial \phi}{\partial \mathbf{n}}\right|_\Gamma = 0
\]
The applied voltages were set across channel side walls ($x=10$ for channel 1 and $x=0$ for channel 2).

The electric potential distribution solved from the Poisson equation is shown in Figure \ref{mesh}(c, d). 
As expected, the field profiles around each pore are symmetric. The electric field is strongest inside the nanopores and decays rapidly away from the pore opening.  The electric field distribution in the common space above the pores (Fig.~\ref{mesh}c) is well approximated by a dipole distribution as each nanopore is approximated by a point-like current source \cite{small2018}.

The time evolution of the probability distribution is governed by the convective diffusion equation following the default Neumann boundary conditions. 
The timestep was 0.5\,ms with a total simulation time of 0.21\,s for TOF and 0.23\,s for resensing. 
The updated probability distribution was calculated at each timestep using an Euler time scheme. 
The cumulative escape probability was calculated via the volume integral of the probability distribution over the relevant volume (channel 2 for TOF, common chamber plus channel 2 for resensing).

To assess uncertainty in the fitting process, bootstrap resampling was applied. 
A total of 20 resampled datasets, with the same size as the original dataset, were generated with replacement, and the standard deviations of the best-fit parameters from these datasets were used to estimate the error bars.

\subsection{On the suitability of an isotropic, rigid rod diffusion model when interpreting the time-of-flight of between nanopores.}

In this section we elaborate on why flexibility and rotational dynamics can be ignored when modeling the diffusion of our nanostructures between nanopores.

\textbf{Flexibility}: The three single seeds (compact, looped, and fringed-looped) all have a length, $L\approx$\, 70\,nm, much shorter than the persistence length of a similarly structured DNA nanotube~\cite{pl}, $L_p\sim$\,2.5$\times$10$^4$\,nm, so modeling them as rigid rods is not controversial. 
To assess the appropriateness of the rigid rod model for the nunchuck seed dimer, we must consider the bend angle distribution of its dsDNA linker and how much that bending affects the diffusion of the dimer.

The variance in the end-to-end bend angle of a dsDNA, $\sigma^2$, is determined by the ratio of its length, $\ell$, to the persistence length of dsDNA, $L_p\sim$\,150\,bp, according to~\cite{cai2020}
\begin{equation*}
\sigma^2 = 2\ell/L_p\,.   
\end{equation*}
Therefore, the bend angle of a 32-bp dsDNA linker in our nunchuck seeds is expected to be normally distributed about 0$^\circ$ with a variance of 0.43 radians, corresponding to a standard deviation of $<$~40$^\circ$.  
This expectation is consistent with experimental observation~\cite{Cai2020DNA}.

Our nunchuck seeds have a total length, $L\sim$\,150~nm, which is much larger than their diameter $D\sim$\,10~nm.
In this long-rod limit, the translational diffusion coefficient, $D_t$, of a rod with a single bend half-way along its length is expected to be no more than 10\% smaller than that of the same rod unbent, for any bend angle~\cite{wegener1981}.
For bend angles $<$~45$^\circ$, the reduction in $D_t$ is $<$~5\%.

\textbf{Rotational dynamics:} The orientational autocorrelation function of a rotationally diffusing object is
\begin{equation*}
    \langle\mathbf{u}(t)\cdot \mathbf{u}(0)\rangle =\mathrm{exp}(-2D_r t)
\end{equation*}
where $\mathbf{u}$ is a unit vector perpendicular to the axis of rotation, $D_r$ is the rotational diffusion constant about that axis, and the rotational decorrelation time is $\tau _r= 1/(2D_r)$.

For a rod-like particle, rotation about the axis parallel to the rod is so fast as to be negligible in water at room temperature.  Consequently, we focus on the decorrelation time for rotation perpendicular to the long axis, given by:
\begin{equation*}
    D_r =\frac{3k_B T(\ln{(L/b)-\gamma)}}{\pi \eta _s L^3}
\end{equation*}
where $\eta _s =$\,0.89\,mPa·s is the solution viscosity, $T=$\,298\,K is the temperature, $b$ is the rod diameter (12\,nm for our seeds), $L$ is the rod length, and $\gamma =$\,0.8 is a correction factor~\cite{doi1986theory}.

For the compact single seeds, which have a length $L\approx$\, 70\,nm, we estimate $D_r \approx$\,1.24$\times$10$^4$\,rad$^2/$s.
This corresponds to a rotational decorrelation time, $\tau_r =$\,0.04\,ms, that is much shorter than their typical time-of-flight (TOF) $\sim$\,27\,ms in our dual-pore system.
For the nunchuck seeds, which are approximately twice as long ($L\approx$\,150\,nm), the estimated $D_r=$\,2.3$\times$10$^3$\,rad$^2/$s, and the decorrelation time $\tau_r =$\,0.2\,ms is still much shorter than their typical TOF$\sim$\,32\,ms.
Note that this estimate approximates the nunchuck seeds as a straight rod.  
For a hinged rod of the same net length, $\tau_r$ will increase, but the effect is less than a factor of 4, and therefore still much shorter than the typical TOF.~\cite{delatorre94}

In summary, the effect of bending is sufficiently small, and the average time-of-flight is sufficeintly long that an isotropic, rigid-rod diffusion model should provide a reasonable estimate of nanostructure length based on our TOF data.

\clearpage

\section{Multi-translocation events analysis}
\subsection{Characteristics of translocation events are similar for both pores}
\begin{figure}[h!]
    \centering
    \includegraphics[width=0.9\linewidth]{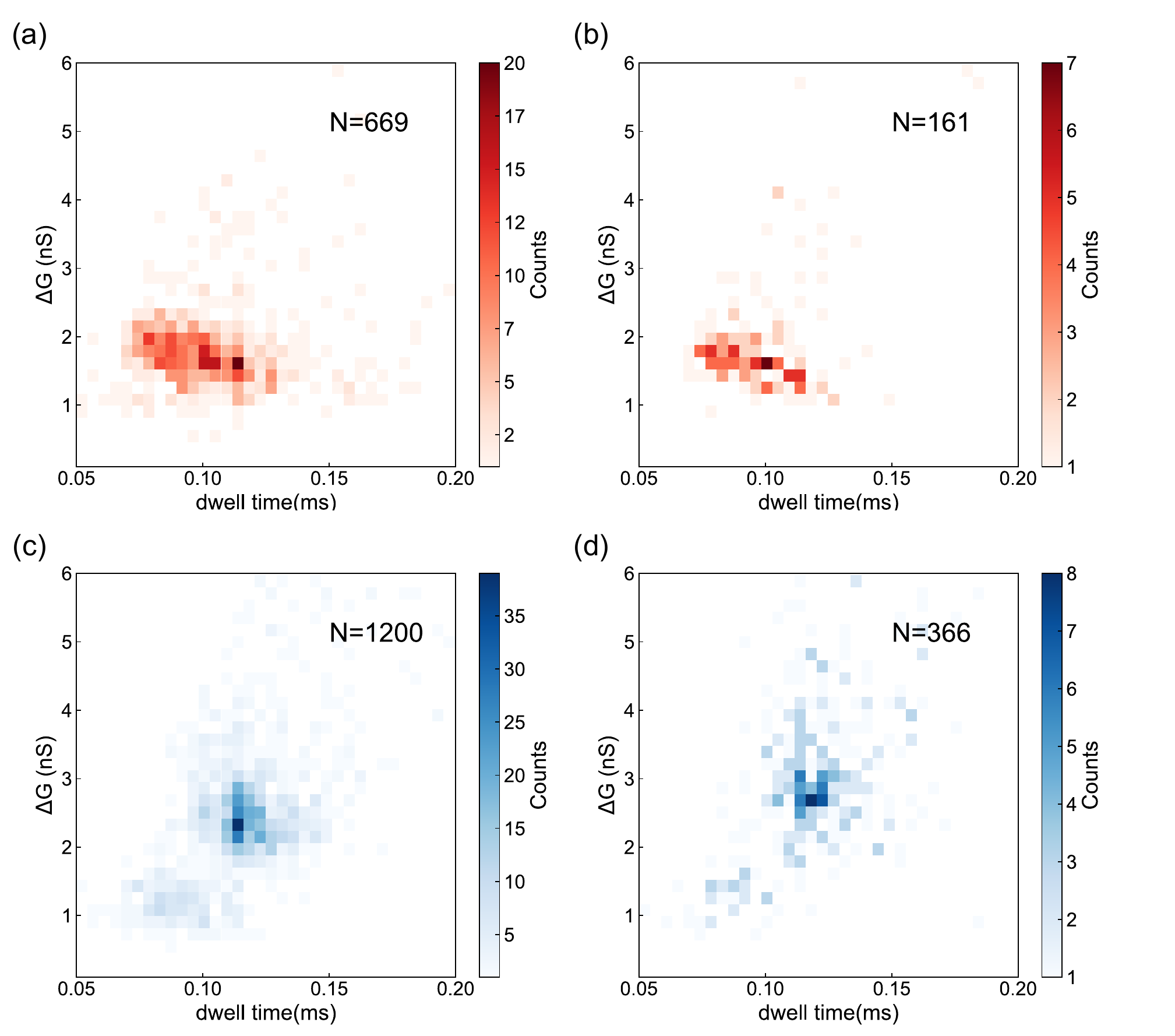} 
  \caption{Density heatmaps of the conductance blockade versus dwell time for all translocation events through pore 1 (a,c) and pore 2 (b,d), for compact seed monomers (a,b - red tones) and nunchuck seeds (c,d - blue tones).
 }
  \label{pore1pore2}
\end{figure}
\newpage

\subsection{Distributions are more localized after averaging over multiple translocation events for each single nanostructure}
\begin{figure}[h!]
    \centering
    \includegraphics[width=0.5\linewidth]{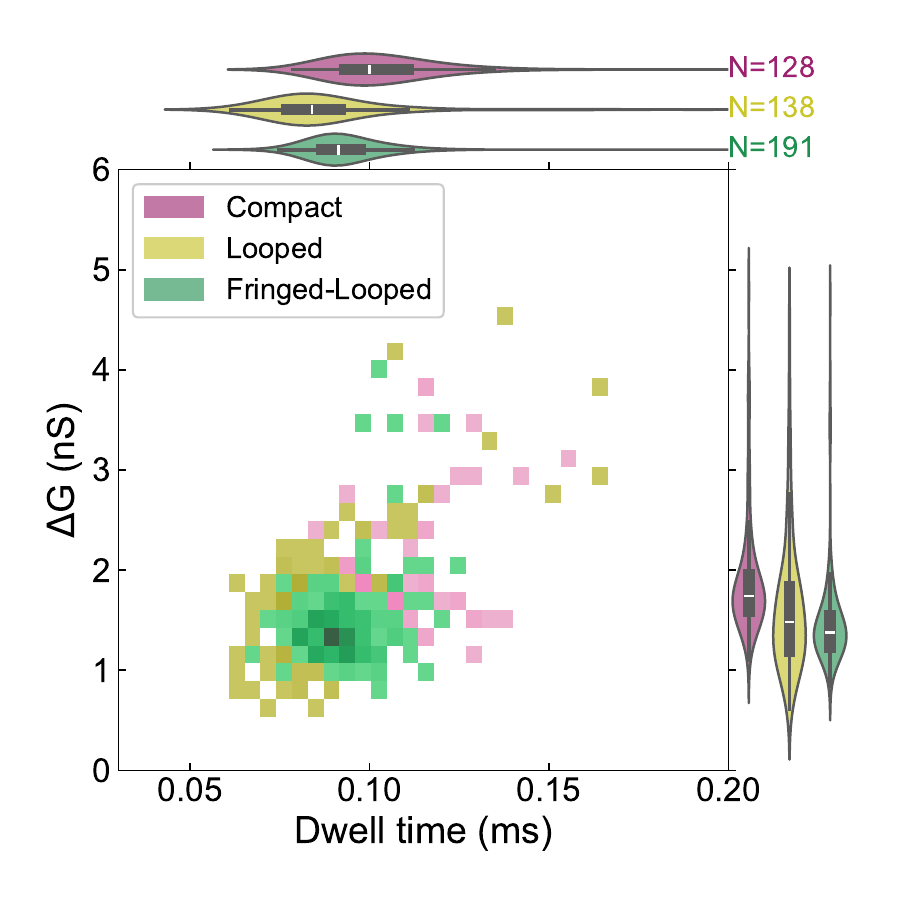} 
  \caption{Density heatmap of mean conductance blockade and dwell time over multi-translocation events for compact, looped and fringed-looped samples, with corresponding marginal violin plots. Only constructs achieving RN 3 or greater are included.
 }
  \label{average}
\end{figure}

\subsection{Nanostructure escape during resensing}
\begin{figure}[h!]
    \centering
    \includegraphics[width=1\linewidth]{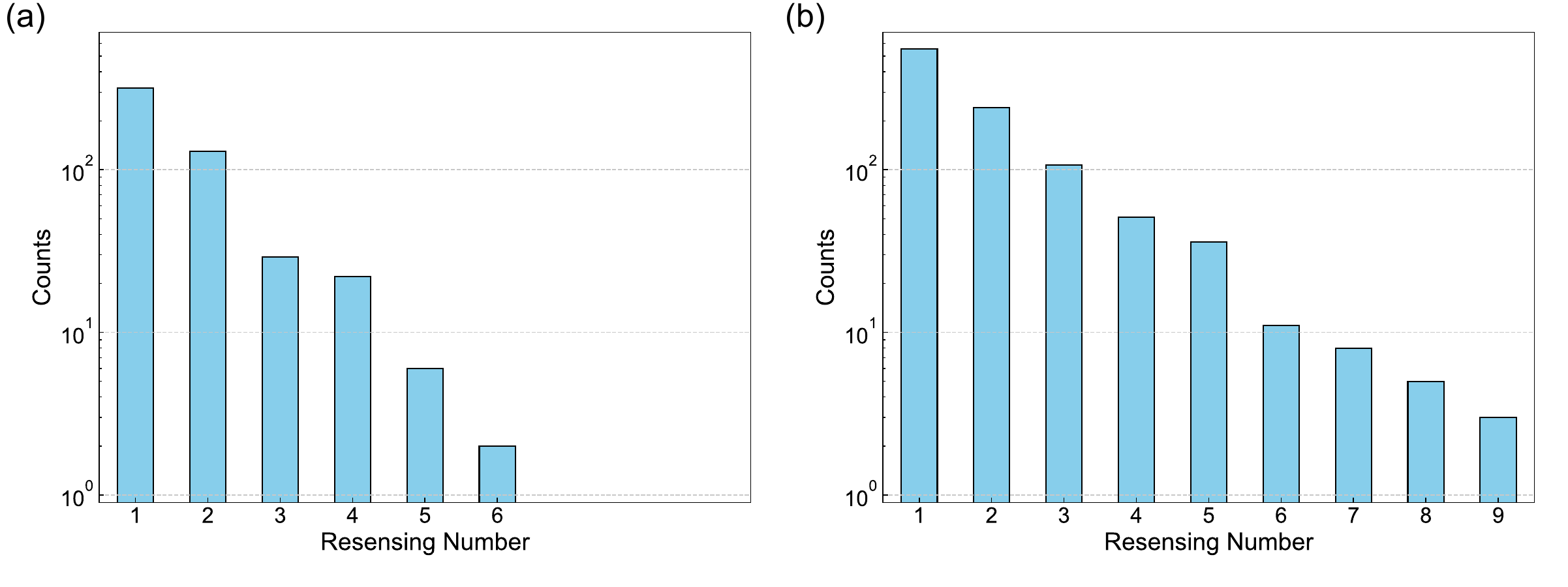} 
  \caption{The number of observed events for each resensing number (RN) for compact seed monomers (a) and nunchuck seeds (b), starting from RN 1 which represents pore 1 resensing. 
  (RN 0 (first pore 1 translocation) events are not shown because only RN 0 events that were followed by an RN 1 event were included in our analysis.) 
  A portion of both nanostructures escaped between each resensing event, but nunchuck seeds exhibited a lower escape rate than compact seed monomers. 
 }
  \label{escape}
\end{figure}
\newpage
\section{DNA nanostructures self-assembly}
\subsection{Recipes}
\begin{table}[h!]
\centering

\begin{tabular}{| l | l | l | l |}
\hline
compact seed monomers & \makecell{Concentration\\ needed (nM)} & \makecell{Concentration\\ at (nM)} & Amount to add (ul) \\
\hline
H2O &  &  & 8.35 \\
\hline
10xTAE 0.125M Mg & 1x& 10x& 5.00 \\
\hline
P3024 Staples & 200 & 2780 & 3.60 \\
\hline
P3024 & 20& 327.5 & 3.05 \\
\hline
SEs core adapters& 200 & 1000 & 10.00 \\
\hline
SEs adapter strand No.2& 400 & 2000 & 10.00 \\
\hline
seed 1 loop staples & 200 & 2000 & 5.00 \\
\hline
fringe blocker & 200 & 2000 & 5.00 \\
\hline
total &  &  & 50.00 \\
\hline
\end{tabular}
\end{table}

\begin{table}[h!]
\centering

\begin{tabular}{| l  |l  |l  |l  |}
\hline
looped seeds &  \makecell{Concentration\\ needed (nM)} & \makecell{Concentration\\ at (nM)} & Amount to add (ul) \\
\hline
H2O &  &  & 13.35 \\
\hline
10xTAE 0.125M Mg & 1x& 10x& 5.00 \\
\hline
P3024 Staples& 200 & 2780 & 3.60 \\
\hline
P3024 & 20& 327.5 & 3.05 \\
\hline
SEs core adapters& 200 & 1000 & 10.00 \\
\hline
SEs adapter strand No.2& 400 & 2000 & 10.00 \\
\hline
 fringe blocker& 200 & 2000 & 5.00 \\
\hline
total &  &  & 50.00 \\
\hline

\end{tabular}

\end{table}

\begin{table}[h!]
\centering

\begin{tabular}{| l  |l  |l  |l  |}
\hline
fringed-looped cylinders &  \makecell{Concentration\\ needed (nM)} & \makecell{Concentration\\ at (nM)} & Amount to add (ul) \\
\hline
H2O &  &  & 38.35 \\
\hline
10xTAE 0.125M Mg & 1x& 10x& 5.00 \\
\hline
P3024 Staples& 200 & 2780 & 3.60 \\
\hline
P3024 & 20& 327.5 & 3.05 \\
\hline
total &  &  & 50.00 \\
\hline

\end{tabular}

\end{table}

\newpage
\subsection{Protocol}
\begin{figure}[h!]
    \centering
    \includegraphics[width=0.5\linewidth]{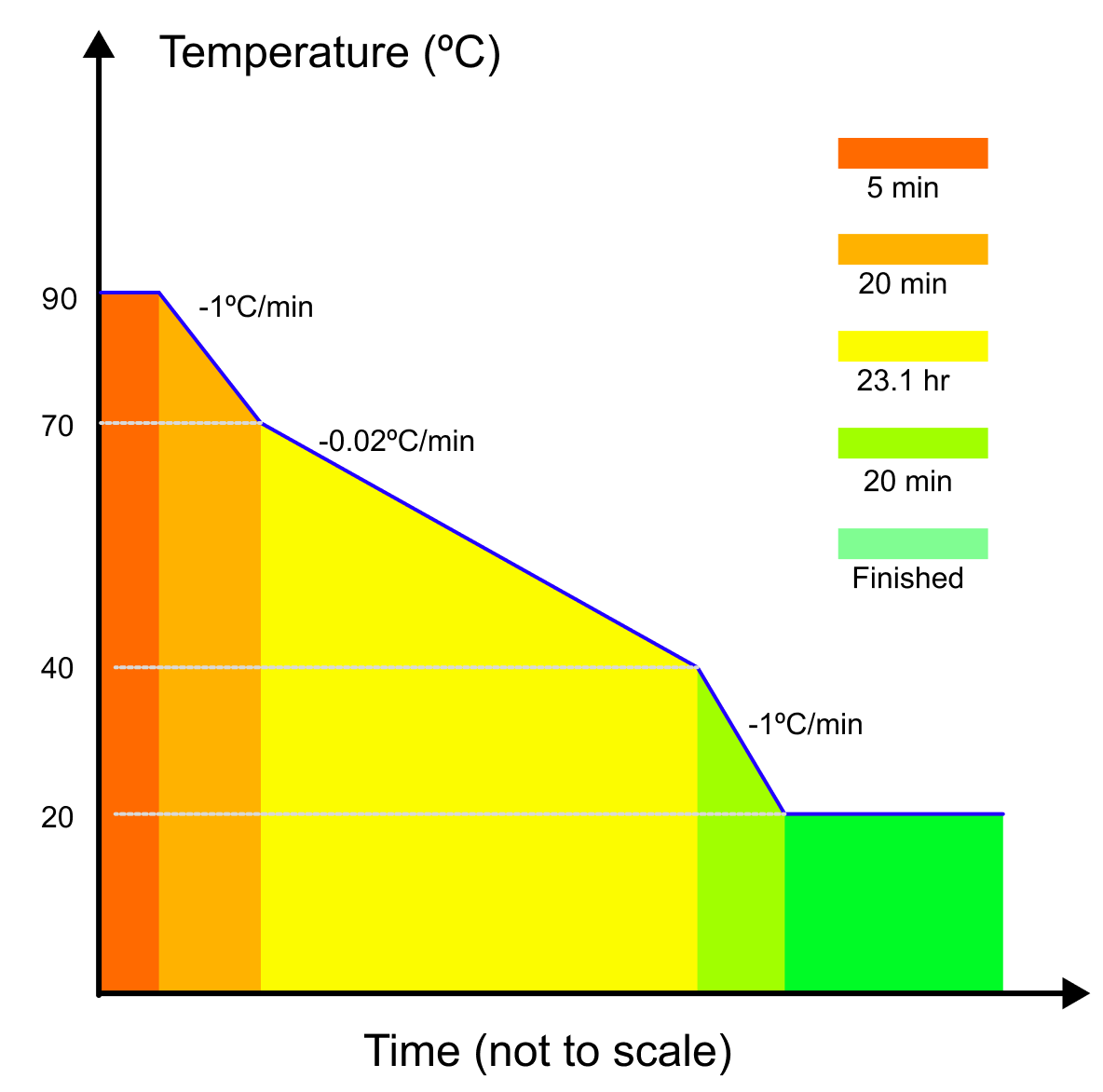}
    \caption{Temperature annealing profile for nanostructure self-assembly.}
    \label{fig:anneal}
\end{figure}

Fringed-looped cylinders, looped seeds and compact seed monomers were self-assembled by thermal annealing (Mastercycler, Eppendorf) following a 24h temperature profile (Fig.~\ref{fig:anneal}).
Nunchuck seeds were made by, first, separately annealing compact seed monomers (Fig.~\ref{fig:anneal}), each with one strand of the double-stranded linker (Section S7.4), mixing the annealed solutions 1:1 and allowing them to hybridize at 31$^\circ$C for 24 hours. 
The hybridized linkers were then ligated (cat no. M0202, New England Biolabs) for 2 hours at room temperature and the ligase was subsequently deactivated by addition of proteinase K (cat no. P8107, New England Biolabs)\cite{Cai2020DNA}. 
The ligation products were gel-purified alongside the three other nanostructures (Fig. S2). 

\newpage
\section{DNA strand sequences}
\subsection{p3024 scaffold}
\tt{\seqsplit{
AATAGTGGACTCTTGTTCCAAACTGGAACAACACTCAACCCTATCTCGGGCTATTCTTTTGATTTATAAGGGATTTTGCCGATTTCGGGGTACCTACGAAGAGTTCCAGCAGGGATTCCAAGAAATGGCCAATGAAGATTGGATCACCTTTCGCACTAAGACCTACTTGTTTGAGGAGTTTCTGATGAATTGGCACGACCGCCTCAGGAAAGTGGAGGAGCATTCTGTGATGACTGTCAAGCTCCAATCTGAGGTGGACAAATATAAGATTGTTATCCCTATCCTGAAGTACGTCCGCGGAGAACACCTGTCACCCGATCACTGGCTGGATCTGTTCCGCTTGCTGGGTCTGCCTCGCGGCACATCTCTGGAGAAACTGCTGTTCGGTGACCTGCTGAGAGTTGCCGATACCATCGTGGCCAAGGCTGCTGACCTGAAAGATCTGAACTCACGCGCCCAGGGTGAAGTGACCATCCGCGAAGCACTCAGGGAACTGGATTTGTGGGGCGTGGGTGCTGTGTTCACACTGATCGACTATGAGGACTCCCAGAGCCGCACCATGAAGCTGATCAAGGATTGGAAGGACATCGTCAACCAGGTGGGCGACAATAGATGCCTCCTGCAGTCCTTGAAGGACTCACCATACTATAAAGGCTTTGAAGACAAGGTCAGCATCTGGGAAAGGAAACTCGCCGAACTGGACGAATATTTGCAGAACCTCAACCATATTCAGAGAAAGTGGGTTTACCTCGAACCAATCTTTGGTCGCGGAGCCCTGCCCAAAGAGCAGACCAGATTCAACAGGGTGGATGAAGATTTCCGCAGCATCATGACAGATATCAAGAAGGACAATCGCGTCACAACCTTGACTACCCACGCAGGCATTCGCAACTCACTGCTGACCATCCTGGACCAATTGCAGAGATGCCAGCGCAGCCTCAACGAGTTCCTGGAGGAGAAGCGCAGCGCCTTCCCTCGCTTCTACTTCATCGGAGACGATGACCTGCTGGAGATCTTGGGCCAGTCAACCAATCCATCCGTGATTCAGTCTCACCTCAAGAAGCTGTTTGCTGGTATCAACTCTGTCTGTTTCGATGAGAAGTCTAAGCACATTACTGCAATGAAGTCCTTGGAGGGAGAAGTTGTGCCATTCAAGAATAAGGTGCCCTTGTCCAATAACGTCGAAACCTGGCTGAACGATCTGGCCCTGAGATGAAGAAGACCCTGGAGCAGCTGCTGAAGGAGTGCGTGACAACCGGACGCAGCTCTCAGGGAGCTGTGGACCCTTCTCTGTTCCCATCACAGATCCTGTGCTTGGCCGAACAGATCAAGTTTACCGAAGATGTGGAGAACGCAATTAAAGATCACTCCCTGCACCAGATTGAGACACAGCTGGTGAACAAATTGGAGCAGTATACTAACATCGACACATCTTCCGAGGACCCAGGTAACACAGAGTCCGGTATTCTGGAGCTGAAACTGAAAGCACTGATTCTCGACATTATCCATAACATCGACGTGGTCAAGCAGCTGAACCAAATCCAAGTGCACACCACCGAAGATTGGGCCTGGAAGAAGCAGTTGAGGTTCTACATGAAGTCCGACCACACCTGTTGCGTTCAGATGGTTGACAGCGAGTTCCAGTACACCTATGAGTACCAAGGAAATGCCAGCAAGCTCGTTTACACTCCACTCACTGACAAGTGTTACCTCACCTTGACACAGGCTATGAAGATGGGCCTGGGAGGCAACCCATACGGTCCAGCTGGCACTGGTAAGACAGAGAGCGTTAAGGCACTCGGAGGTCTGCTGGGCAGGCAGGTCCTCGTGTTCAACTGTGATGAAGGAATCGACGTTAAGTCCATGGGAAGAATCTTTGTTGGCCTCGTTAAGTGTGGAGCTTGGGGTTGCTTCGACGAGTTCAACAGGCTGGAGGAATCTGTGCTGAGCGCCGTCTCTATGCAGATCCAGACCATCCAGGACGCATTGAAGAACCACAGGACCGTCTGCGAGCTGTTGGGTAAGGAAGTGGAGGTGAACTCCAACTCCGGAATCTTCATCACAATGAATCCCGCAGGTAAAGGATATGGAGGAAGACAGAAACTCCCAGACAACCTGAAGCAGCTGTTCCGCCCAGTGGCTATGTCCCATCCAGACAATGAGCTGATCGCCGAAGTCATCCTCTATTCCGAGGGATTCAAAGATGCTAAAGTTCTCTCCAGAAAGCTCGTGGCCATCTTCAATCTGTCAAGAGAACTCCTGACACCTCAGCAGCATTACGACTGGGGTCTGAGAGCCCTCAAGACCGTCCTGAGAGGTTCAGGAAATCTCCTCAGGCAGCTGAACAAGAGCGGTACAACACAGAATGCAAATGAGAGCCACATTGTCGTCCAGGCTCTGAGGCTGAATACCATGTCAAAGTTCACATTCACAGACTGCACAAGATTTGACGCTCTGATTAAAGATGTGTTCCCTGGTATTGAACTCAAAGAAGTGGAGTATGACGAGCTGAGCGCCGCTTTGAAGCAGGTGTTTGAGGAGGCTAACTATGAGATTATCCCTAATCAGATCAAGAAAGCATTGGAACTGTATGAACAGCTGTGTCAGAGGATGGGAGTGGTGATTGTGGGCCCATCAGGCGCAGGTAAGAGCACTCTCTGGAGAATGCTGAGAGCAGCACTGTGCAAGACTGGAAAGGTGGTGAAGCAATACACCATGAATCCCGGATCCACGCGCCCTGTAGCGGCGCATTAAGCGCGGCGGGTGTGGTGGTTACGCGCAGCGTGACCGCTACACTTGCCAGCGCCCTAGCGCCCGCTCCTTTCGCTTTCTTCCCTTCCTTTCTCGCCACGTTCGCCGGCTTTCCCCGTCAAGCTCTAAATCGGGGGCTCCCTTTAGGGTTCCGATTTAGTGCTTTACGGCACCTCGACCCCAAAAAACTTGATTTGGGTGATGGTTCACGTAGTGGGCCATCGCCCTGATAGACGGTTTTTCGCCCTTTGACGTTGGAGTCCACGTTCTTT}}
\normalfont
%
\newpage
\subsection{SEs adapters}

\textnormal
Note that the *\_2 labeled strands are referred to as No.2 strands, and *\_1, *\_3, and *\_5 strands are referred to as core strands in our nanostructure making recipe (section S6).

\begin{table}[h!]
    \centering
    \begin{tabular}{|l|l|}
        \hline
        \textbf{Name} & \textbf{Sequence} \\ \hline
        SEs\_1\_1&          \tt{ACCACGGTAGGAGTGCTTGTCGGACGAGTTGTGTTACCTGGGTCC}\\ \hline
        SEs\_1-2\_3&          \tt{GACAAGCACTCCTGTCCTGATGCGACTGAGCTGGACTCGTCC}\\ \hline
        SEs\_2\_5&          \tt{CCAGAATACCGGACTCCCAGCTCAGTCGCATCAGGACCTCCATGC}\\ \hline
        SEs\_3\_1&          \tt{GACAAGGTCTTCAGCGTCATGCTCGCTACAGCTTGCTGGCATTTC}\\ \hline
        SEs\_3-4\_3&          \tt{CATGACGCTGAAGCGAACGGAAGCACGGTTGCCTGTAGCGAG}\\ \hline
        SEs\_4\_5&          \tt{GAGTGGAGTGTAAACGAGGCAACCGTGCTTCCGTTCGAGTCGGCA}\\ \hline
        SEs\_5\_1&          \tt{GAAGGCTGTAGGTTCGTGCTCAGAGACCTAAGCTCCACACTTAAC}\\ \hline
        SEs\_5-6\_3&          \tt{GAGCACGAACCTAGTGGTGATCGCTCGTCGCCAGAGGTCTCT}\\ \hline
        SEs\_6\_5&          \tt{TCGTCGAAGCAACCCCCTGGCGACGAGCGATCACCACACTCCAAG}\\ \hline
        SEs\_7\_1&          \tt{ACAGCATGAGGTCACGTCTACGTCCACTCCAGCTGCTTCAGGTTG}\\ \hline
        SEs\_7-8\_3&          \tt{GTAGACGTGACCTGGTTCGATGCGAGACTTCGTGGAGTGGAC}\\ \hline
        SEs\_8\_5&          \tt{TAGCCACTGGGCGGAACACGAAGTCTCGCATCGAACCGCTATCCG}\\ \hline
        SEs\_9\_1&          \tt{CCGTGCCTGCCATAGGACACCACCATGAGCTCTTGTTCAGCTGCC}\\ \hline
        SEs\_9-10\_3&          \tt{GGTGTCCTATGGCGATGGAAGTCGAGGTGACTCGCTCATGGT}\\ \hline
        SEs\_10\_5&          \tt{ATTCTGTGTTGTACCGCGAGTCACCTCGACTTCCATCACGCCTTG}\\ \hline
        SEs\_11\_1&          \tt{GAGCATCCTCACACTGCGAGTAGCAAGGCTCTTGATCTGATTAGG}\\ \hline
        SEs\_11-12\_3&          \tt{ACTCGCAGTGTGACCGTACCTCTGTCCTAGTCGTGCCTTGCT}\\ \hline
        SEs\_12\_5&          \tt{TACAGTTCCAATGCTTACGACTAGGACAGAGGTACGGATCAGCCA}\\ \hline
        1-2\_SEs\_2&          \tt{GTCTGGCATGGAGACCGTGGTAGGTA}\\ \hline
        3-4\_SEs\_2&          \tt{GTCTGTGCCGACTACCTTGTCAGGTA}\\ \hline
        5-6\_SEs\_2&          \tt{GTCTGCTTGGAGTCAGCCTTCAGGTA}\\ \hline
        7-8\_SEs\_2&          \tt{GTCTGCGGATAGCCATGCTGTAGGTA}\\ \hline
        9-10\_SEs\_2&          \tt{GTCTGCAAGGCGTAGGCACGGAGGTA}\\ \hline
        11-12\_SEs\_2&          \tt{GTCTGTGGCTGATGGATGCTCAGGTA}\\ \hline
    \end{tabular}
 \label{tab:SEs_seq}
\end{table}

\newpage
\subsection{REs adapters}

\textnormal
Note that the *\_2 labeled strands are referred to as No.2 strands, and *\_1, *\_3, and *\_5 strands are referred to as core strands in our nanostructure making recipe (section S6).

\begin{table}[h!]
    \centering
    \begin{tabular}{|l|l|}
        \hline
        \textbf{Name} & \textbf{Sequence} \\ \hline
        REs\_1\_1&          \tt{CAGCCAAGACGCAGGTAGCGAGACAGAGCTGTGTTACCTGGGTCC}\\ \hline
        REs\_1-2\_3&          \tt{TCGCTACCTGCGTTCGTCGGATGGTGAGGTCCACGCTCTGTC}\\ \hline
        REs\_2\_5&          \tt{CCAGAATACCGGACTCGTGGACCTCACCATCCGACGACACGAGCA}\\ \hline
        REs\_3\_1&          \tt{CACGGAGTCGAAGCGTAGGACGGTAGCCAAGCTTGCTGGCATTTC}\\ \hline
        REs\_3-4\_3&          \tt{GTCCTACGCTTCGGACCTTGGTGATGCTGGACTGTGGCTACC}\\ \hline
        REs\_4\_5&          \tt{GAGTGGAGTGTAAACGCAGTCCAGCATCACCAAGGTCGCTCGGCA}\\ \hline
        REs\_5\_1&          \tt{CAGAGCCACGGCATGGTCTTGCGTTGGAGAAGCTCCACACTTAAC}\\ \hline
        REs\_5-6\_3&          \tt{CAAGACCATGCCGACCTCATCCTCGCTTTCGGTGCTCCAACG}\\ \hline
        REs\_6\_5&          \tt{TCGTCGAAGCAACCCCCACCGAAAGCGAGGATGAGGTGCGGACGA}\\ \hline
        REs\_7\_1&          \tt{GGAAGCGTCTACGGCAGTGACCGATCTCCCAGCTGCTTCAGGTTG}\\ \hline
        REs\_7-8\_3&          \tt{GTCACTGCCGTAGCTCACGAGGCACAACCACAGCGGAGATCG}\\ \hline
        REs\_8\_5&          \tt{TAGCCACTGGGCGGAAGCTGTGGTTGTGCCTCGTGAGTCGCAGGA}\\ \hline
        REs\_9\_1&          \tt{CAACCGTCGTTCCACAGGACTCGCACTTCCTCTTGTTCAGCTGCC}\\ \hline
        REs\_9-10\_3&          \tt{AGTCCTGTGGAACACCACGAGACGCCATCGAGCGGAAGTGCG}\\ \hline
        REs\_10\_5&          \tt{ATTCTGTGTTGTACCGCGCTCGATGGCGTCTCGTGGTAAGCCTGA}\\ \hline
        REs\_11\_1&          \tt{CAGCGTCGCAAGGCTACGAGTCAGACAGGTCTTGATCTGATTAGG}\\ \hline
        REs\_11-12\_3&          \tt{ACTCGTAGCCTTGGACCGCACTCACCACTGCTCGCCTGTCTG}\\ \hline
        REs\_12\_5&          \tt{TACAGTTCCAATGCTTCGAGCAGTGGTGAGTGCGGTCATCCGTGC}\\ \hline
        1-2\_REs\_2&          \tt{GACCATGCTCGTGCTTGGCTGGCATT}\\ \hline
        3-4\_REs\_2&          \tt{GACCATGCCGAGCACTCCGTGGCATT}\\ \hline
        5-6\_REs\_2&          \tt{GACCATCGTCCGCTGGCTCTGGCATT}\\ \hline
        7-8\_REs\_2&          \tt{GACCATCCTGCGAACGCTTCCGCATT}\\ \hline
        9-10\_REs\_2&          \tt{GACCATCAGGCTTGACGGTTGGCATT}\\ \hline
        11-12\_REs\_2&          \tt{GACCAGCACGGATCGACGCTGGCATT}\\ \hline
    \end{tabular}
 \label{tab:REs_seq}
\end{table}
\newpage

\subsection{p3024 seed staples}
\begin{table}[h!]
\label{tabel: p3024 seed stp part 1}
\begin{filecontents*}{p3024_Staple_seq.csv}
Name,Sequence
stp\_h1t1,CCGAAATCGCGTCCGGCGTCCTGCTTTTGCAGGACGTTTTGTCACGCATCTCCA
stp\_h1t3,GAATAGCCAACAGAGAGCCAGGACTTTTGTCCTGGCTTAGGGTCCACGACGTTA
stp\_h1t5,AGTTTGGATCTGTTCGAGTGCCTGTTTTCAGGCACTTTGCCAAGCAAATGGCAC
stp\_h1t7,TGCTGCTCAATTGCGTGCCATCGCTTTTGCGATGGCTTTCTCCACATCTTCGGT
stp\_h1t9,TACCTGCGTGTGTCTCCCGTGGCGTTTTCGCCACGGTTAATCTGGTTTCAGCTG
stp\_h1t11,CCCATCCTTGTTAGTACGGAACCGTTTTCGGTTCCGTTTACTGCTCAATGTCGA
stp\_h2t2,CTGAATCATCGTTCAGGTGGCAGCTTTTGCTGCCACTTCCAGGTTTCAGCTCCC
stp\_h2t4,TTGATACCGGGCACCTGGCTCCGCTTTTGCGGAGCCTTTATTCTTGCAGGATCT
stp\_h2t6,TGTGCTTACCTCCAAGCACCGCTGTTTTCAGCGGTGTTGACTTCATCCCAATCT
stp\_h2t8,AGAACCTCGTGCACTTGGCACCTGTTTTCAGGTGCCTTGGATTTGGGCAGGGAG
stp\_h2t10,CTGAACGCCGTCGATGCCTGCCTGTTTTCAGGCAGGTTTTATGGATCAATTTGT
stp\_h2t12,TCATAGGTGCTTTCAGGCCAGTGCTTTTGCACTGGCTTTTTCAGCTTCGGAAGA
stp\_h3t1,GGGCCAGACGGATGGACTCCTGCGTTTTCGCAGGAGTTTTGGTTGACGATGAAG
stp\_h3t3,TTGGACAAAGCAAACACAGGAGCGTTTTCGCTCCTGTTGCTTCTTGTTCTCCTC
stp\_h3t5,AACTTCTCGACTTCTCGTGCTGCGTTTTCGCAGCACTTATCGAAACGCATCTCT
stp\_h3t7,TCGGTGGTAACTGCTTACGCTGGCTTTTGCCAGCGTTTCTTCCAGGTGCAGTAA
stp\_h3t9,CTTGACCAAACAGGTGCGGACCTGTTTTCAGGTCCGTTTGGTCGGACGTATGGG
stp\_h3t11,GAATCAGTGTACTGGAGTGGTCCGTTTTCGGACCACTTACTCGCTGGCCTGTGT
stp\_h4t2,TGCGGAAAAGGGAAGGCGTGGAGCTTTTGCTCCACGTTCGCTGCGCAGGTGAGA
stp\_h4t4,GCGATTGTCGTTGAGGCGGAGCACTTTTGTGCTCCGTTCTGCGCTGAGACAGAG
stp\_h4t6,CGAATGCCTCCAGGATCCTACCGCTTTTGCGGTAGGTTGGTCAGCAAACGCTCT
stp\_h4t8,GCCCAGCACCAGTGCCCTGGACGCTTTTGCGTCCAGTTAGCTGGACCTTCATGT
stp\_h4t10,CCTTCATCCAGGCCCAGCTGACGCTTTTGCGTCAGCTTTCTTCATATCAACCAT
stp\_h4t12,CAAAGATTGGTAACACCGGATCGCTTTTGCGATCCGTTTTGTCAGTCTTGGTAC
stp\_h5t1,TAGAAGCGTCTTCATCGGTGGTCGTTTTCGACCACCTTCACCCTGTCGACCAAA
stp\_h5t3,CAGGAACTCCTTCTTGGCTCGCTGTTTTCAGCGAGCTTATATCTGTTCTCTGAA
stp\_h5t5,GCAATTGGTGCGTGGGGTCTGGCGTTTTCGCCAGACTTTAGTCAAGCGTCCAGT
stp\_h5t7,CTGTCTTAGACCTCCGTCGGAGGCTTTTGCCTCCGATTAGTGCCTTGTGAGTTG
stp\_h5t9,TTGCCTCCACAGTTGACCGAGGCGTTTTCGCCTCGGTTACACGAGGGTCCTGGA
stp\_h5t11,CAAGGTGACTTCCCATGCTCGGACTTTTGTCCGAGCTTGGACTTAACGCTCAGC
stp\_h6t2,ACCTGGTTCGAGGTAACTGGCTCGTTTTCGAGCCAGTTACCCACTTCATGATGC
stp\_h6t4,TCAAGGACAGGTTCTGTCGGAGCGTTTTCGCTCCGATTCAAATATTGTTGTGAC
stp\_h6t6,GTCTTCAATTTCCTTTGACCTGGCTTTTGCCAGGTCTTCCCAGATGAGCTCGCA
stp\_h6t8,TTCACCTCTGTGGTTCCCACCTCGTTTTCGAGGTGGTTTTCAATGCACCTGCCT
stp\_h6t10,GATTCATTATCTGCATGCCGAGCGTTTTCGCTCGGCTTAGAGACGGCGTCGATT
stp\_h6t12,TTTCTGTCCTCCAGCCCGACGGACTTTTGTCCGTCGTTTGTTGAACGAGGCCAA
stp\_h7t1,GATTGGTTGACGATGTACCAGCCGTTTTCGGCTGGTTTCCTTCCAAGTCCTCAT
stp\_h7t3,TATGGTTGTGCAGGAGCGGAGGCGTTTTCGCCTCCGTTGCATCTATCCACGCCC
stp\_h7t5,TCGGCGAGAGCCTTTAGCCTAGCGTTTTCGCTAGGCTTTAGTATGGTCGCGGAT
stp\_h7t7,GACGGTCCCACTTCCTGTGAGGCGTTTTCGCCTCACTTTACCCAACCTGACCTT
stp\_h7t9,TGGTCTGGGTGATGAACTCGGTCGTTTTCGACCGAGTTGATTCCGGGCATCTTT
stp\_h7t11,ACAGATTCTTCCTCCACGGCAGGCTTTTGCCTGCCGTTTATCCTTTTTCGGCGA
stp\_h8t2,TTTCTCCAAGTGTGAACTGGTGGCTTTTGCCACCAGTTCACAGCACTGTCGCCC
stp\_h8t4,TCGGCAACCAGTTCCCCACGACGCTTTTGCGTCGTGTTTGAGTGCTTGAGTCCT
stp\_h8t6,CTTTCAGGCACCCTGGTCTGGTCGTTTTCGACCAGATTGCGCGTGAAGATGGCC
stp\_h8t8,TCAGGAGTTCTGGAGAACCAGGCGTTTTCGCCTGGTTTGAACTTTAAGTTGGAG
stp\_h8t10,TCTCAGACCGGAATAGTGGCGAGCTTTTGCTCGCCATTAGGATGACACCTGCGG
stp\_h8t12,TTTCCTGATTGTCTGGGTCGCTGCTTTTGCAGCGACTTATGGGACATCTGGGAG
stp\_h9t1,AGTCGATCGAGATGTGCTCGCTGCTTTTGCAGCGAGTTCCGCGAGGGATCGGGT
stp\_h9t3,CACAAATCTCTCAGCACCGTCCGCTTTTGCGGACGGTTGGTCACCGTTCAGGAT
stp\_h9t5,GGTCACTTTCAGCAGCCGAGGTCGTTTTCGACCTCGTTCTTGGCCACACCTCAG
stp\_h9t7,ACGAGCTTTCTCTTGAGTCAGGCGTTTTCGCCTGACTTCAGATTGAGTTCAGAT
stp\_h9t9,GAATCCCTCCCAGTCGCTCTGGACTTTTGTCCAGAGTTTAATGCTGTGAATGTG
stp\_h9t11,TCAGCTCAACCTCTCACGAGGCACTTTTGTGCCTCGTTGGACGGTCAGAGCCTG
\end{filecontents*}
\csvreader[my table=Predefined table]
{p3024_Staple_seq.csv}{}
{\name & \seq}
\end{table} 

\begin{table}[h!]
\centering
\label{tabel: p3024 seed stp part 2}
\begin{adjustbox}{max width=\textwidth} 
\begin{tabular}{| l | l |}
\hline
\textbf{Name} & \textbf{Sequence} \\
\hline
stp\_h10t2& GAAAGGTGTTCTCCGCTCGGCTCGTTTTCGAGCCGATTGGACGTACAACAGCAG \\
\hline
stp\_h10t4& CAGAAACTCAATCTTACCGCAGGCTTTTGCCTGCGGTTTATTTGTCCGATGGTA \\
\hline
stp\_h10t6& TCCACTTTTTGACAGTCGCTCGTGTTTTCACGAGCGTTCATCACAGTTTAATCA \\
\hline
stp\_h10t8& GAGTTCAAAATCTTGTGTCGGTGCTTTTGCACCGACTTGCAGTCTGCTGAGGTG \\
\hline
stp\_h10t10& GCGGCGCTCATGGTATCCGAGGACTTTTGTCCTCGGTTTCAGCCTCTTGAGGGC \\
\hline
stp\_h10t12& CATAGTTATGTGGCTCCGCTTGGCTTTTGCCAAGCGTTTCATTTGCTGAGGAGA \\
\hline
stp\_h11t1& GACAGGTGATCCAATCTGCGGTCGTTTTCGACCGCATTTTCATTGGTAGGTACC \\
\hline
stp\_h11t3& AGGGATAACCTCAAACCTGCGTCGTTTTCGACGCAGTTAAGTAGGTAATCAAAA \\
\hline
stp\_h11t5& ATTGGAGCCCTGAGGCCACAGGCGTTTTCGCCTGTGTTGGTCGTGCGTTGTTCC \\
\hline
stp\_h11t7& GAGCGTCATACCAGGGCGCTTGCGTTTTCGCAAGCGTTAACACATCAATGCTCC \\
\hline
stp\_h11t9& AACTTTGACAGCTCGTCTCGGCACTTTTGTGCCGAGTTCATACTCCAGTGCTCT \\
\hline
stp\_h11t11& GACGACAAGCCTCCTCGTCCACCGTTTTCGGTGGACTTAAACACCTTCACCACT \\
\hline
stp\_h12t2& TGAGAGCTGGCAAAATCGGATGGCTTTTGCCATCCGTTCCCTTATACTTAGTGC \\
\hline
stp\_h12t4& GTGATGGGCGAGATAGGGCATCCGTTTTCGGATGCCTTGGTTGAGTCAATTCAT \\
\hline
stp\_h12t6& AAACTTGAACAAGAGTTGGCTGCGTTTTCGCAGCCATTCCACTATTTTGCACAG \\
\hline
stp\_h12t8& TGATCTTTTCAGCATTCGGTCGGCTTTTGCCGACCGTTCTCCAGAGACTTCTTT \\
\hline
stp\_h12t10& TCACCAGCCCTGATGGGCAGTCGCTTTTGCGACTGCTTGCCCACAAGCTTCAAA \\
\hline
stp\_h12t12& TGTGTCGACTGACACACGGTGCCGTTTTCGGCACCGTTGCTGTTCAGATAATCT \\
\hline

\end{tabular}
\end{adjustbox}
\end{table}

\newpage
\subsection{32bp nunchuck linker}
Linker strands were either purchased with $5^\prime$-phosphates or were $5^\prime$-phosphorylated in house using T4 polynucleotide kinase (cat no. M0201, New England Biolabs).
\\
\\
\noindent
L1: \texttt{ATGCTATTGGCCCCGTGTCTCCTTTGGGCAGGGCTCCGTGAATCTGGTCTGCTCTTTTGGAGACACGG}	\\		
\textnormal
L2: \texttt{GGCCAATAGCATGTGATCGGTGTTTGGGCAGGGCTCCGTGAATCTGGTCTGCTCTTTTCACCGATCAC}

\subsection{Loop staples}
\begin{table}[h!]
    \centering
    \begin{tabular}{|l|l|}
        \hline
        \textbf{Name} & \textbf{Sequence} \\ \hline
        LS\_1& \tt{GCTGCGCGTGATTCATGGTGT}\\ \hline
        LS\_2& \tt{TGCTTCACCACGACTCCGCGAAAGAGGGAGC}\\ \hline
        LS\_3& \tt{GTAAAGCTTAGAGCAGGAAGGGAAGAAAAACGTCAGCGCGTG}\\ \hline
        LS\_4& \tt{AAAAACCGCGAGAATTGACGGGGAAAGCGGGGTCGAACCATCAC}\\ \hline
            LS\_5& \tt{GCGAACGTGGTCTATCAGGGC}\\\hline
            LS\_6& \tt{GATCCGGAACCACCCGGAACCCTCGGTCAC}\\\hline
             LS\_7& \tt{CCCCGATACTAAATACACCCGCCGCGCTTAATGTGAGGTGCC}\\\hline
            LS\_8& \tt{GCTGGCAAGTGTAGAAGAGCGGGCGCTAGGGC}\\\hline
             LS\_9& \tt{GGCCCACTACGCGCCGCTACAGGAAGGGCG}\\\hline
            
    \end{tabular}
\end{table}
\subsection{Fringe blockers}
\begin{table}[h!]
    \centering
    \begin{tabular}{|l|l|}
        \hline
        \textbf{Name} & \textbf{Sequence} \\ \hline
        FringeBlocker\_1-2& \tt{TTTTGCTCCAGGGTCTTCTTCACTCCTTCAGCAGCTTTTT}\\ \hline
        FringeBlocker\_3-4& \tt{TTTTAGCAGGTCATCGTCTCCTGGCCCAAGATCTCCTTTT}\\ \hline
        FringeBlocker\_7-8& \tt{TTTTTGGTGCGGCTCTGGGATCCTTGATCAGCTTCATTTT}\\ \hline
        FringeBlocker\_9-10& \tt{TTTTAACAGATCCAGCCAGTCAGACCCAGCAAGCGGTTTT}\\ \hline
    \end{tabular}
 
\end{table}

\section{Chuck design}
\begin{figure}[h!]
    \centering
    \includegraphics[width=0.6\linewidth]{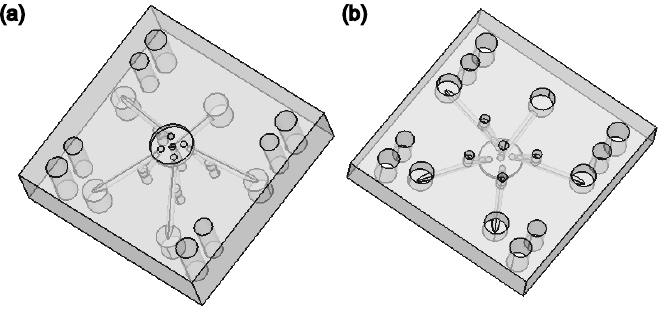} 
  \caption{3D view of the design for the microfluidic chuck. Dual-nanopore chips are mounted onto the pit on the bottom surface (a). Samples and buffer are loaded through ports on the top surface (b). 
 }
  \label{escape}
\end{figure}
\newpage

\section{SEM image of the dual-nanopore chip}
\begin{figure}[h!]
    \centering
    \includegraphics[width=0.8\linewidth]{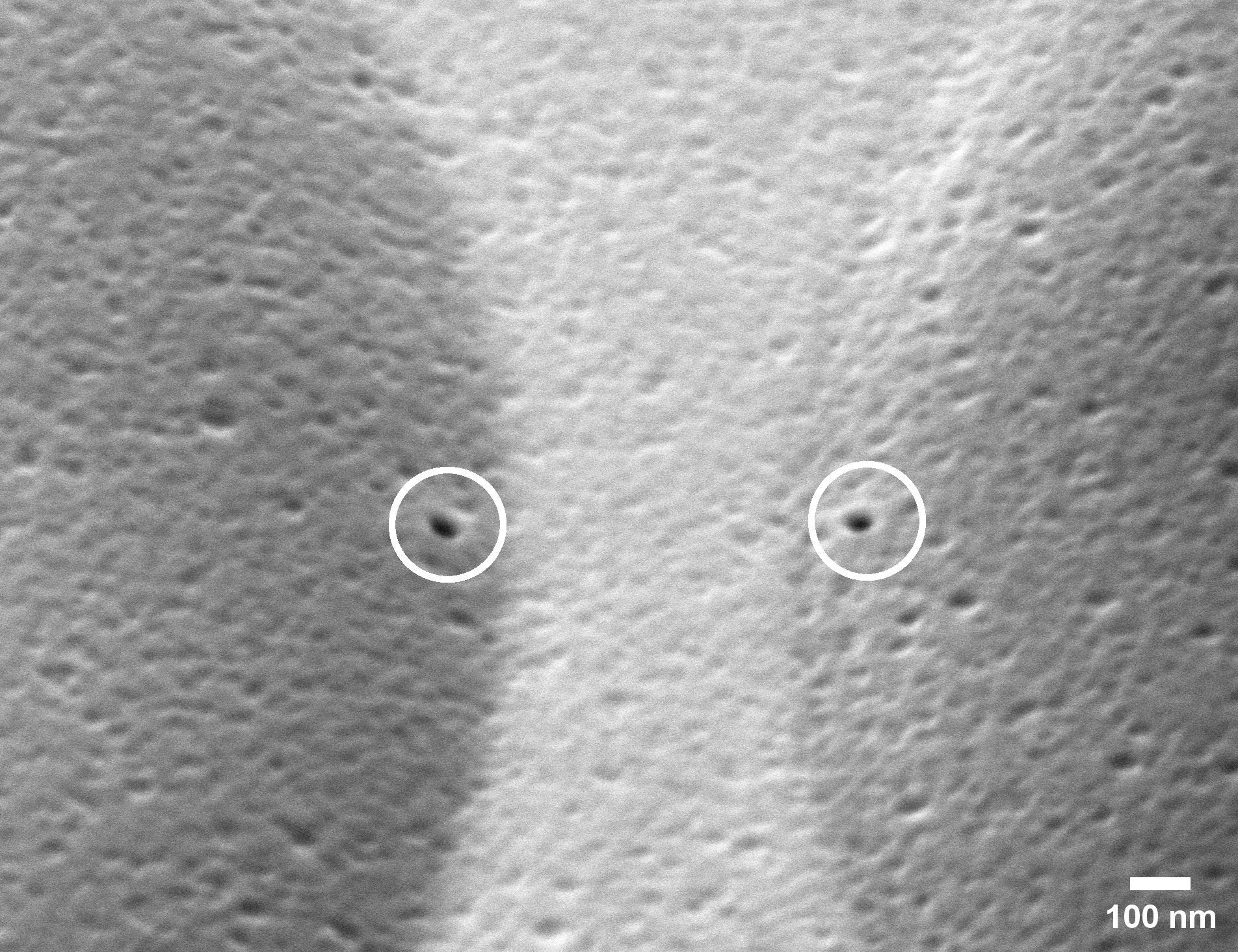} 
  \caption{SEM image of the nanopore chip used in the experiments. The two nanopores are highlighted in the white circles.
 }
  \label{SEM}
\end{figure}
\newpage

FIB drilling can reliably produce pores of controlled diameters $>$25\,nm. 
We found nanopores $\sim$30\,nm in diameter were optimal for this study: small enough to provide signals with high signal-to-noise ratio and large enough to reduce clogging risk.
In a typical batch, the nanopores have an average diameter of 32.6\,nm with a standard deviation of 5.3\,nm, as measured from an IV curve taken at the beginning of each experiment. 
The rate of functional dual-pore chips was 
$>$60\%.

\bibliographystyle{plain}
\bibliography{reference}